\documentclass{pasa}
\usepackage{rotating}
\usepackage{graphicx}
\usepackage{multirow}
\usepackage{relsize}

\title[CRAFT Sensitivity]{The Performance and Calibration of the CRAFT Fly's Eye Fast Radio Burst Survey}

\author[C.\ W.\ James et al.]{C.\ W.\ James$^{1,2}$\thanks{clancy.james@curtin.edu.au}, K.\ W.\ Bannister$^3$, J.-P.\ Macquart$^{1,2}$, R.\ D.\ Ekers$^{1,3}$, S.\ Oslowski$^{4}$, R.\ M.\ Shannon$^{1,2,4}$, J.\ R.\ Allison$^{5}$, A.\ P.\ Chippendale$^{3}$, J.\ D.\ Collier$^{3,6}$, T.\ Franzen$^{7}$, A.\ W.\ Hotan$^{7}$, M.\ Leach$^{3}$, D.\ McConnell$^{3}$, M.\ A.\ Pilawa$^{3}$, M.\ A.\ Voronkov$^{3}$, M.\ T.\ Whiting$^{3}$
\affil{$^1$International Centre for Radio Astronomy Research, Curtin University, Bentley, WA 6102, Australia}
\affil{$^2$ARC Centre of Excellence for All-sky Astrophysics (CAASTRO), Australia}
\affil{$^3$Australia Telescope National Facility, CSIRO Astronomy and Space Science, PO Box 76, Epping, NSW 1710, Australia}
\affil{$^4$Swinburne University of Technology, PO Box 218, Hawthorn, VIC 3122, Australia}
\affil{$^5$Sub-Dept. of Astrophysics, Department of Physics, University of Oxford, Denys Wilkinson Building, Keble Rd., Oxford, OX1 3RH, UK}
\affil{$^6$School of Computing, Engineering, and Mathematics, Western Sydney University, Locked
Bag 1797, Penrith, NSW 2751, Australia}
\affil{$7$CSIRO Astronomy and Space Science, Australia Telescope National Facility, PO Box 1130,
Bentley WA 6102, Australia}
}

\jid{PASA}
\doi{10.1017/pas.\the\year.xxx}
\jyear{\the\year}

\usepackage{aas_macros}
\usepackage{hyperref} 
\hypersetup{colorlinks,citecolor=blue,linkcolor=blue,urlcolor=blue}

\begin{document}

\begin{frontmatter}
\maketitle

\begin{abstract}
Since January 2017, the Commensal Real-time ASKAP Fast Transients survey (CRAFT) has been utilising commissioning antennas of the Australian SKA Pathfinder (ASKAP) to survey for fast radio bursts (FRBs) in fly's eye mode. This is the first extensive astronomical survey using phased array feeds (PAFs). A total of 23 FRBs have been reported --- here, we present a calculation of the sensitivity and total exposure of the survey that detected the first 20 of these bursts, using the pulsars B1641-45 (J1644-4559) and B0833-45 (J0835-4510, i.e.\ Vela) as calibrators. The design of the survey allows us to benchmark effects due to PAF beamshape, antenna-dependent system noise, radio-frequency interference, and fluctuations during commissioning on timescales from one hour to a year. Observation time, solid-angle, and search efficiency are calculated as a function of FRB fluence threshold. Using this metric, effective survey exposures and sensitivities are calculated as a function of the source counts distribution. Statistical `stat' and systematics `sys' effects are treated separately. The implied FRB rate is significantly lower than the $37$\,sky$^{-1}$\,day$^{-1}$ calculated using nominal exposures and sensitivities for this same sample by \citet{craft_nature}. At the Euclidean power-law index of $-1.5$, the rate is $12.7_{-2.2}^{+3.3}\,{\rm (sys)} \, \pm \, 3.6\,{\rm (stat)}$\,sky$^{-1}$\,day$^{-1}$ above a threshold of $56.6 \pm 6.3\,{\rm (sys)}$\,Jy\,ms, while for the best-fit index for this sample of $-2.2$, it is $20.7_{-1.7}^{+2.1} \,{\rm (sys)}\, \pm 5.8\,{\rm (stat)}$\,sky$^{-1}$\,day$^{-1}$ above a threshold of $40.4 \pm 1.2\,{\rm (sys)}$\,Jy\,ms. This strongly suggests that these calculations be performed for other FRB-hunting experiments, allowing meaningful comparisons to be made between them.
\end{abstract}

\begin{keywords}
methods: data analysis -- telescopes --- surveys
\end{keywords}
\end{frontmatter}

\section{INTRODUCTION}
\label{sec:intro}

Fast radio bursts (FRBs) are enigmatic transient phenomena. First detected as ``A bright millisecond radio burst of extragalactic origin'' by \citet{2007Sci...318..777L}, subsequent observations \citep{2013Sci...341...53T} have established a population of these events occurring at a rate of thousands per sky per day. These bursts are all the more remarkable in that not only are their dispersion measures well in excess of the Galactic contribution, but that few have plausible associations with galaxies in the nearby universe, and only one has had a host galaxy confirmed \citep{2017ApJ...834L...7T}. This makes them intrinsically extremely powerful events, and also suggests their use as cosmological probes. Efforts to study the nature of FRB progenitors and their hosts are ongoing, with a key question being whether or not the repeating FRB \citep{2014ApJ...790..101S,2016Natur.531..202S} is part of the same population.

Even the most basic properties of the FRB population(s) are poorly constrained. Both the estimated rate and spectral index of the cumulative source counts distribution vary greatly with the method used (e.g.\ \citet{2016ApJ...830...75V}), and with each new set of observations  (e.g.\ \citet{2017MNRAS.468.3746C}). The most recent estimate by \citet{2018MNRAS.475.1427B} suggests 800--3200 FRBs per sky per day with fluences above 2\,Jy\,ms, with a cumulative source counts distribution of fluences with power-law index of $-2.2^{+0.6}_{-1.2}$. As pointed out by \citet{2018MNRAS.474.1900M}, however, there are many pitfalls in estimating these parameters, and telescope parameters such as the beam pattern must be extremely well-understood in order to correctly calibrate an FRB survey. Accurate estimation of these effects will become even more important as the sample of detected FRBs is expanded from the initial dominance of Parkes (e.g.\ \citet{{2007Sci...318..777L},2016MNRAS.460L..30C,2018MNRAS.475.1427B}), to ASKAP \citep{craft_nature}, UTMOST \citep{2017PASA...34...45B}, the VLA \citep{0067-0049-236-1-8}, CHIME \citep{2017ApJ...844..161A}, and other instruments with their own unique properties.

The goal of this paper is to develop the methods for such a necessary and detailed calibration, and particularly for the recently published sample of 20 FRBs detected with the ASKAP radio telescope by the CRAFT collaboration \citep{2017ApJ...841L..12B,craft_nature}.

ASKAP, the Australian SKA Pathfinder \citep{2008ExA....22..151J,2009IEEEP..97.1507D,2012SPIE.8444E..2AS,2016SPIE.9906E..2AS}, is an array of 36 $12$\,m antennas located in the Murchison Radio Observatory in Western Australia. It is equipped with Phased Array Feeds (PAFs; \citet{2008RaSc...43.6S04H}), and can simultaneously form 36 beams for a field of view (FoV) of $30$\,deg$^2$ at 1.4\,GHz.
A total of 384 1\,MHz channels between 0.7 and 1.8\,GHz are digitised, with 336 currently available for time-domain analysis.

CRAFT, the Commensal Real-time ASKAP Fast Transients survey \citep{2010PASA...27..272M}, aims to use ASKAP to commensally detect a large number of fast radio bursts in real-time. During the ASKAP commissioning phase CRAFT has been observing using available antennas. Observations have primarily been in fly's eye mode, increasing the field-of-view proportional to the number of observing antennas. From 2017 to early 2018, independent fields at Galactic latitudes of $|b|= 50\pm5^{\circ}$ were targeted, denoted the CRAFT `GL50' survey. The use of a near-constant Galactic latitude avoids any possible latitude-dependence of the FRB rate \citep{2014ApJ...789L..26P,2014ApJ...792...19B,2015MNRAS.451.3278M}, and limits the Galactic contribution to dispersion measure. The CRAFT GL50 survey has now concluded, accumulating a total of 1427 antenna days of data, with 20 FRBs being detected \citep{2017ApJ...841L..12B,craft_nature}. As such, it has accumulated far more FRBs in a stable configuration than any other survey. This both motivates and enables a detailed analysis of ASKAP's sensitivity to FRBs.

As noted by \citet{2018MNRAS.474.1900M}, the FRB detection rate depends on the interaction between antenna beamshape and the observed source counts distribution (the detection rate of FRBs as a function of fluence threshold). With the exception of FRB 121102 \citep{2016Natur.531..202S}, FRBs are poorly localised, and their detected fluence will be related to their true fluence through an unknown factor of the antenna beam pattern. This makes it impossible to reduce survey sensitivity to a characteristic flux/fluence threshold without knowing the relative likelihood of detection at each point in the beam pattern. Rather, the required metric is the survey exposure (area--time product) $E(F_{\rm th})$ as a function of FRB fluence detection threshold $F_{\rm th}$, from which the response to any given hypothesis on the FRB fluence distribution can be calculated.

This paper calculates $E(F_{\rm th})$ for the CRAFT GL50 survey, described in more detail in Section~\ref{sec:obs_and_data} below. Section~\ref{sec:pulsar_cal} describes pulsar calibration observations, which are used to calibrate beam and antenna sensitivities, and account for the effects of radio-frequency interference (RFI) and power fluctuations experienced during commissioning. Section~\ref{sec:beams} describes the use of holographic observations to measure the ASKAP beamshape over all $36$\,beams. Section~\ref{sec:sensitivity} combines these results with an absolute sensitivity calibration to derive $E(F_{\rm th})$, and calculates effective survey parameters under different hypotheses of the FRB integral source counts function. This allows the all-sky FRB rate to be estimated, the implications of which are discussed in Section~\ref{sec:discussion}. Throughout this work, unless otherwise noted, all uncertainties are quoted at the $1\,\sigma$ (68\% confidence) level.

\section{Observations and Data}
\label{sec:obs_and_data}

\subsection{Observation strategy}
\label{sec:obs}

The CRAFT fly's eye observation strategy and data processing pipeline was originally described in detail in \citet{2017ApJ...841L..12B}. Below, the key features are revisited, with some minor updates to the analysis strategy.

CRAFT observations have primarily made use of ASKAP antennas as they became available, with between one and eleven antennas observing simultaneously. Since the beam-formed commensal mode of CRAFT is still being commissioned, antennas have been operating in fly's eye mode, with data from each antenna analysed independently.

\begin{figure*}
\begin{center}
\includegraphics[width=0.49\textwidth]{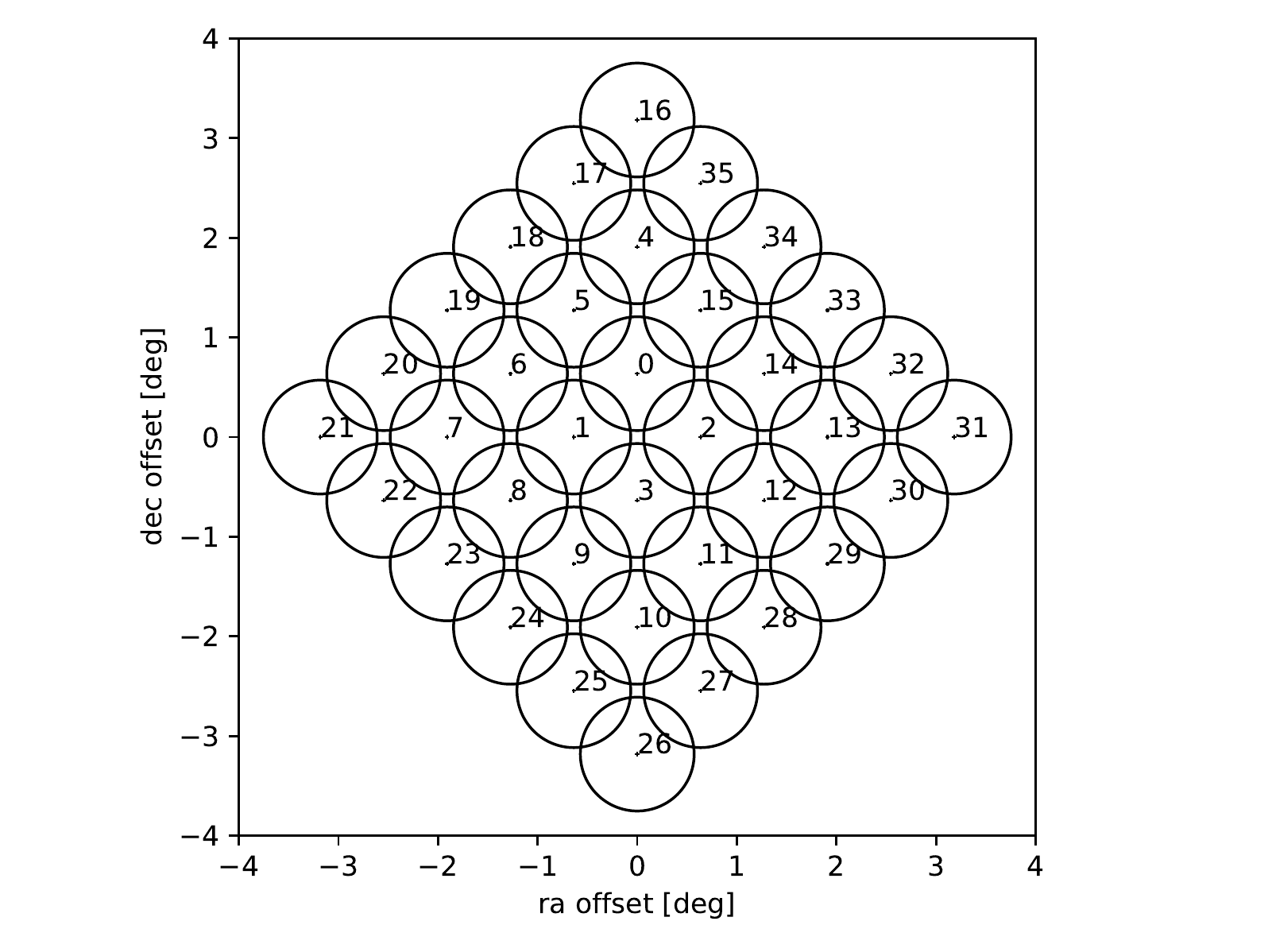} \includegraphics[width=0.49\textwidth]{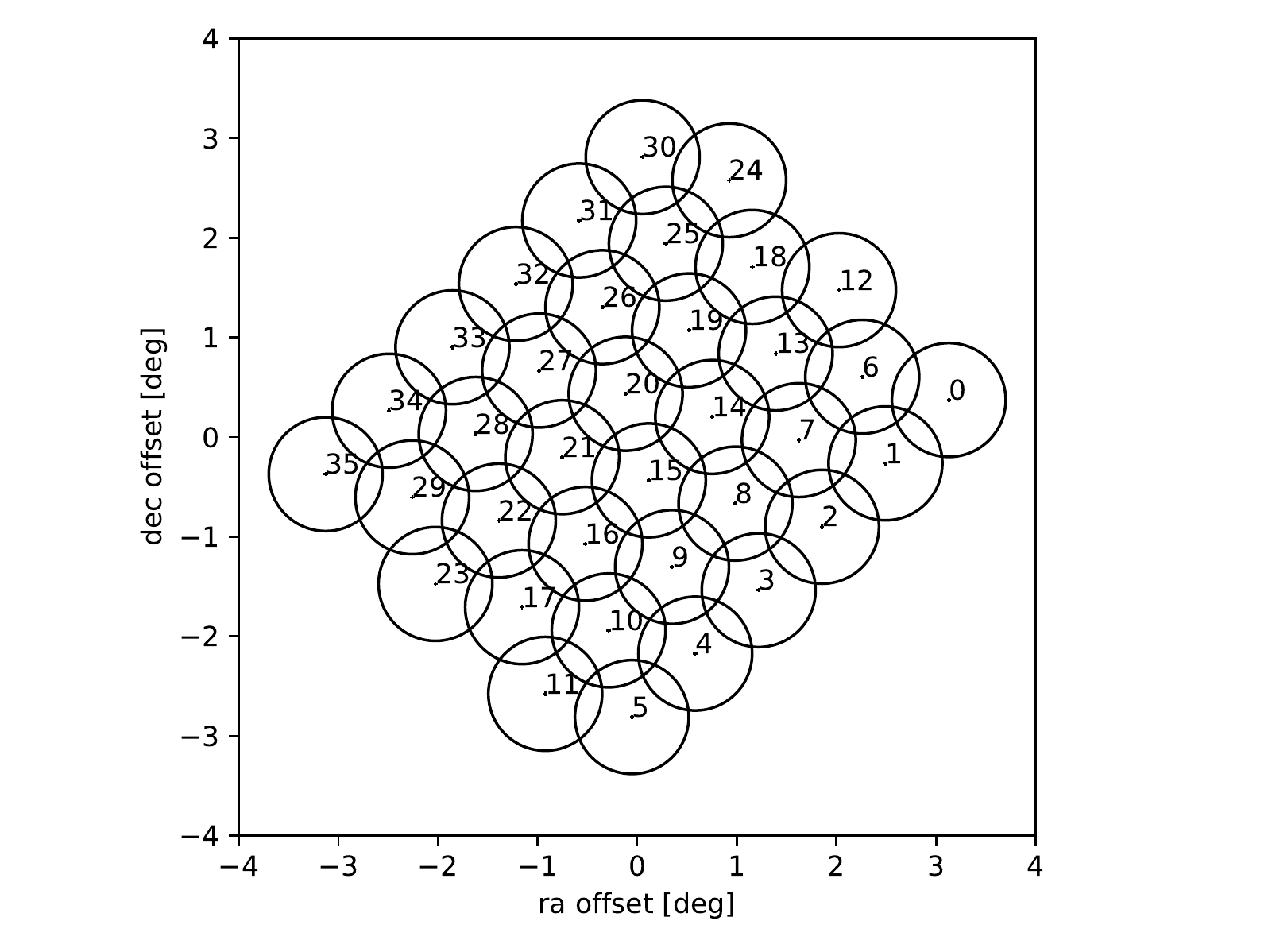}
\caption{Diagrams of ASKAP footprints used in CRAFT FRB searches: `square6x6' (left) and `closepack36' (right), showing beam centre offsets about antenna boresight. In both cases, a pitch angle (angle of separation between beams) of $0.9^{\circ}$ was used. Numbers indicate beam IDs, while the circles indicate the half-power beam width at the central frequency of 1296\,MHz, assuming an Airy beam pattern.} \label{fig:closepack}
\end{center}
\end{figure*}

We have been observing using the `square6x6' footprint prior to March 17th, and the `closepack36' footprint subsequently. The beam patterns of these footprints are shown in Figure~\ref{fig:closepack}. While the overlapping beams reduce the total effective survey area for the closepack36 configuration, they also reduce the importance of sidelobes in rate calculations, and increase the likelihood of a multibeam detection. The process for forming ASKAP beams is described in \citet{2016PASA...33...42M}; this results in minor variations in beam fidelity every time beamforming is performed, while minor variations in gain and phase from each digital receiver port will vary the beamshape with time once beamforming has been performed.

Observations have used a contiguous bandwidth, from $1128$ to $1464$\,MHz, dividing into 336 1\,MHz channels.\footnote{Some very early observations used slightly different frequencies and bandwidths, with negligible contribution to the total survey time.} To reduce the data rate to computationally feasible levels, the squared complex voltages from both polarisations are integrated over 1500 samples, i.e.\ $1.2656$\,ms at ASKAP's 32/27 oversampled rate. This causes dispersion smearing within a channel to exceed the integration time at a DM of 333\,pc\,cm$^{-3}$ at band centre.

These data are then recorded to disk, and searched for FRBs as described below.

\subsection{Data processing and analysis}
\label{sec:data_processing}

CRAFT FRB searches are performed in near-real-time by `FREDDA', a GPU-implementation of the FDMT algorithm \citep{2017ApJ...835...11Z}. It also performs basic checks of data fidelity, such as flagging saturated channels, and subtraction of zero-dispersion artefacts. The search space is restricted to dispersions of between $100$ and $4096$ samples, corresponding to dispersion measures between 95.9 and 3930\,pc\,cm$^{-3}$ in $\sim$0.959\,pc\,cm$^{-3}$ increments. The final stage in FREDDA searches in pulse width space, using $32$ uniform windows over $1,2,3,\ldots,32$ $1.2656$\,ms samples, and returns the candidate with the most significant width.

Metadata on all candidates over $7\,\sigma$ significance are reported by FREDDA. These are then passed to a friends-of-friends algorithm \citep{1982ApJ...257..423H} to merge candidates within two increments in any dimension in search space, i.e.\ DMs within $\pm1.92$\,pc\,cm$^{-3}$ and arrival times within $\pm 1.53$\,ms. The most significant candidate in each group is recorded, and since a great many RFI candidates with widths greater than $16$ samples (20.25\,ms) were found, these are rejected. Remaining candidates above $9.5\,\sigma$ are visually inspected for final confirmation as FRBs. Candidates from each beam are treated independently, although once an FRB is identified, data from neighbouring beams are used for source localisation. Thus, to a good approximation, the dependency of CRAFT sensitivity on FRB arrival direction can be calculated from the sensitivity envelope of all $36$ beams.

The reporting threshold of $7\,\sigma$ for FREDDA candidates was chosen because A: this allows a measurement of pure noise events (observed up to $8\,\sigma$); B: candidates observed above $7\,\sigma$ in two beams can theoretically be excluded as being pure noise events, and identified as FRBs; and C: the distribution in search-parameter space of reported significance about peak significance will be more peaked for true FRBs, aiding in the exclusion of RFI. We have yet to perform a systematic multibeam search as described by B, with only FRB 171216 being coincidentally discovered in this manner \citep{craft_nature}; while C was not required for exclusion of RFI.

\section{Modelling sensitivity and efficiency with pulsar calibration observations}

\label{sec:pulsar_cal}

\begin{figure*}
\begin{center}
\includegraphics[width=0.7\textwidth]{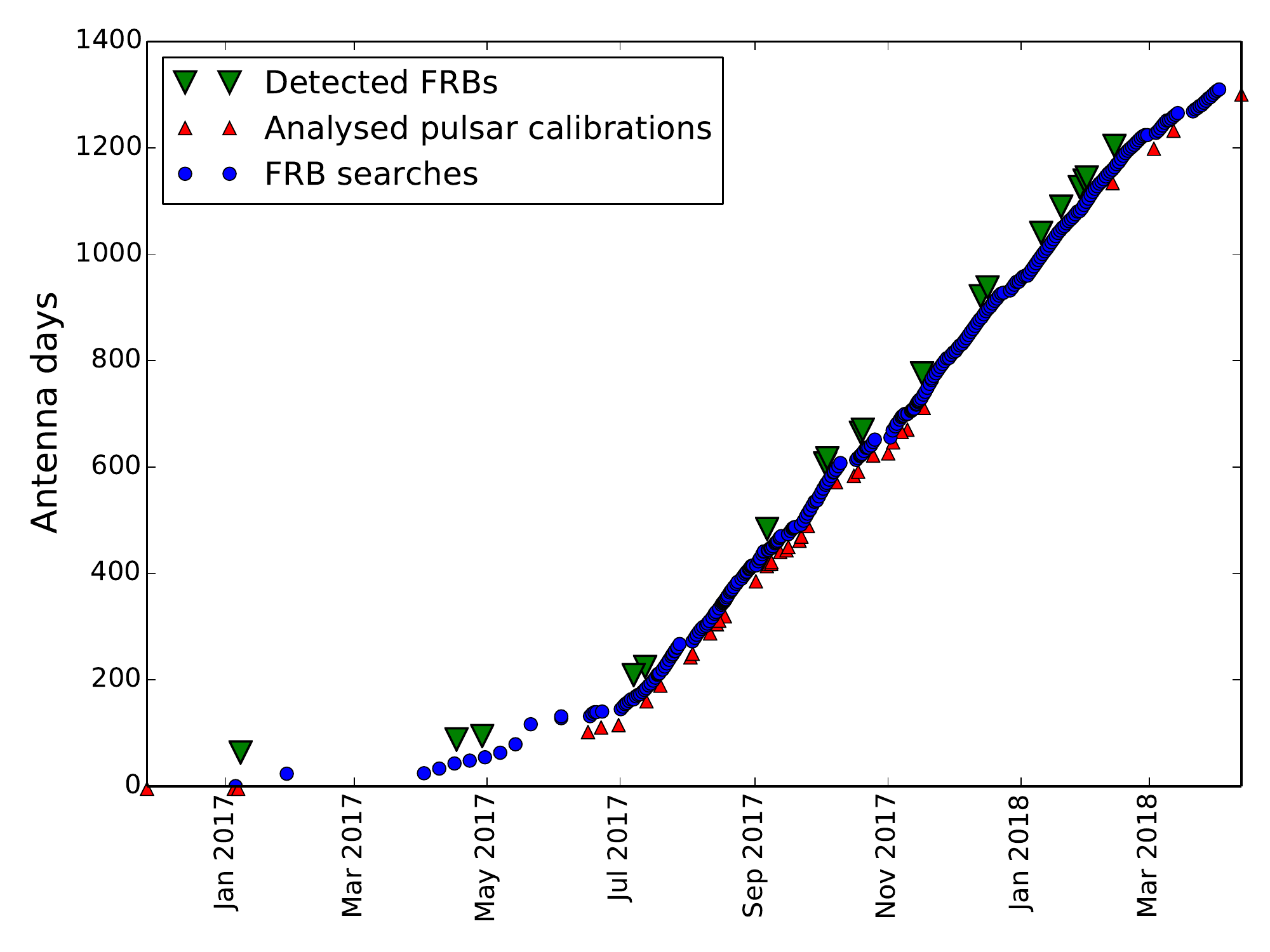}
\caption{Timing of pulsar calibration runs (red triangles) compared to detected FRB times (green inverted triangles) reported in \citet{craft_nature}, and the cumulative FRB search observation time in antenna-days (blue dots). No antenna efficiency factors have been included.}\label{fig:obs_cal_frb}
\end{center}
\end{figure*}

CRAFT observations alternate between 57 minute scans of FRB search fields, with one antenna per field, and 3 minute observations with all antennas on a bright, stable pulsar, either B1641-45 (J1644-4559) or B0833-45 (J0835-4510, Vela). The latter are known as `pulsar check' scans, and are used to verify system performance. Additionally, for each new set of PAF beamformer weights, a `pulsar calibration' observation is performed, in which each beam on all antennas is sequentially pointed at a pulsar, for a period of approximately 100\,s per pointing. Here, we use these pulsar calibration observations to determine the time, antenna, and beam dependence of ASKAP sensitivity. This is then linked to an absolute sensitivity in Section~\ref{sec:absolute}. The effects of the overall ASKAP beam pattern are calculated in Section~\ref{sec:beams}. The timing of these pulsar calibration observations are compared to those of data-taking runs, and detected FRBs, in Figure~\ref{fig:obs_cal_frb}. The analysed calibration runs are determined by the frequency of new beamforming solutions, and the stability of the observing configuration.

\subsection{Fitting method}
\label{sec:fitting_method}

Data from 38 pulsar calibration observations taking during the survey have been analysed. This included all such observations up to November 2017, at which point the observing configuration and fluctuations in sensitivity (see Section~\ref{sec:power_fluctuations}) had stabilised.

The calibration data were processed through FREDDA and friends-of-friends using exactly the same algorithms as for FRB searches, with the exception that candidates down to a DM of $46$\,pc\,cm$^{-3}$ are included. Candidates with DMs within $\pm 4$\,pc\,cm$^{-3}$ of the known values of the target pulsar (67.99\,pc\,cm$^{-3}$ for B0833-45, and 478.8\,pc\,cm$^{-3}$ for B1641-45 \citep{2005AJ....129.1993M} from the on-pulsar beam only are selected. All such candidates from a given calibration observation are binned in terms of their measured signal-to-noise values as determined by FREDDA, $\sigma_{\rm F}$. The estimated fraction of coincidental triggers contaminating this sample is less than 0.1\,\%, as is the loss from pulses with misestimated DMs falling outside the DM search range.

\begin{figure*}
\begin{center}
\includegraphics[width=0.49\textwidth]{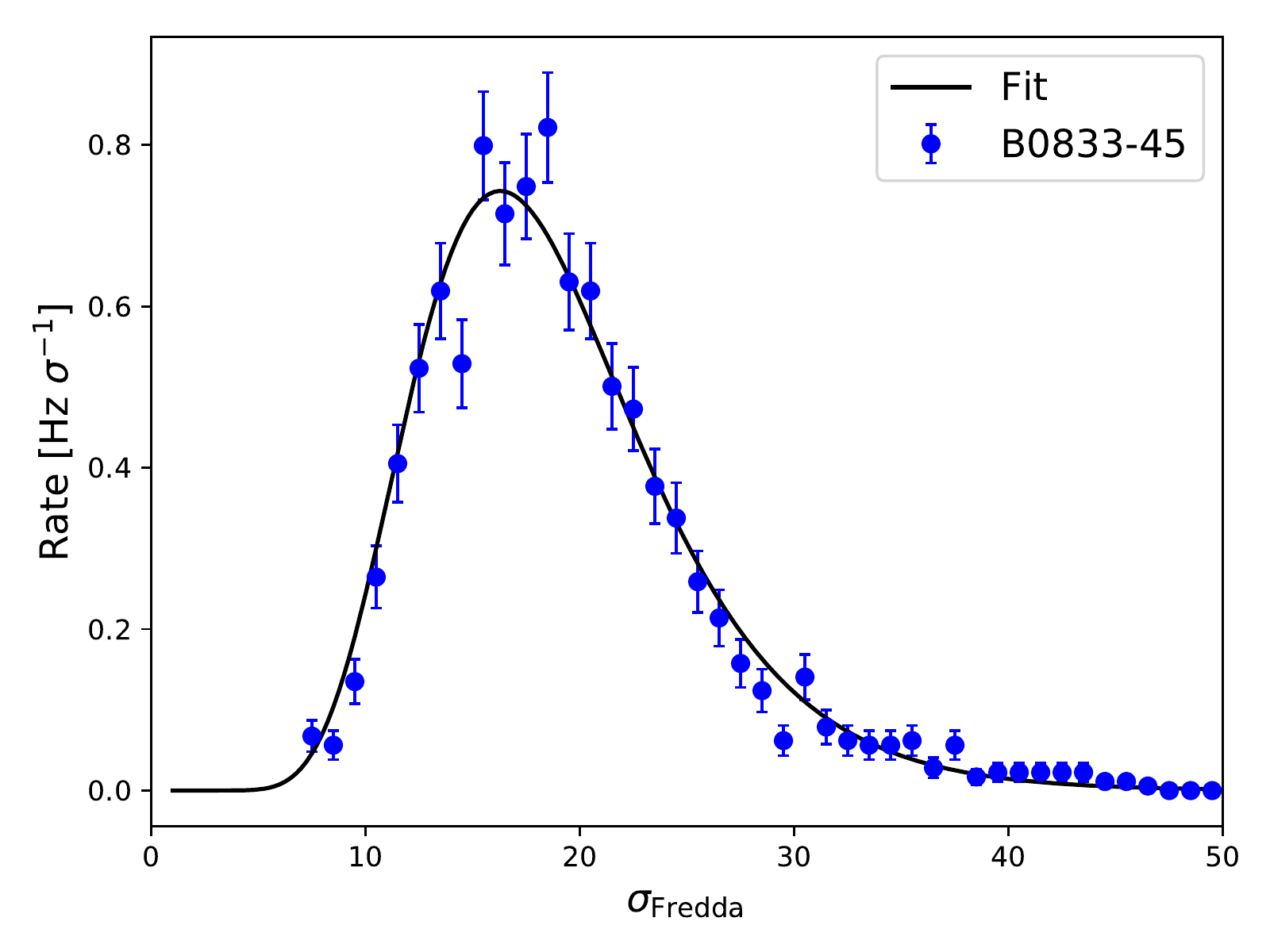} \includegraphics[width=0.49\textwidth]{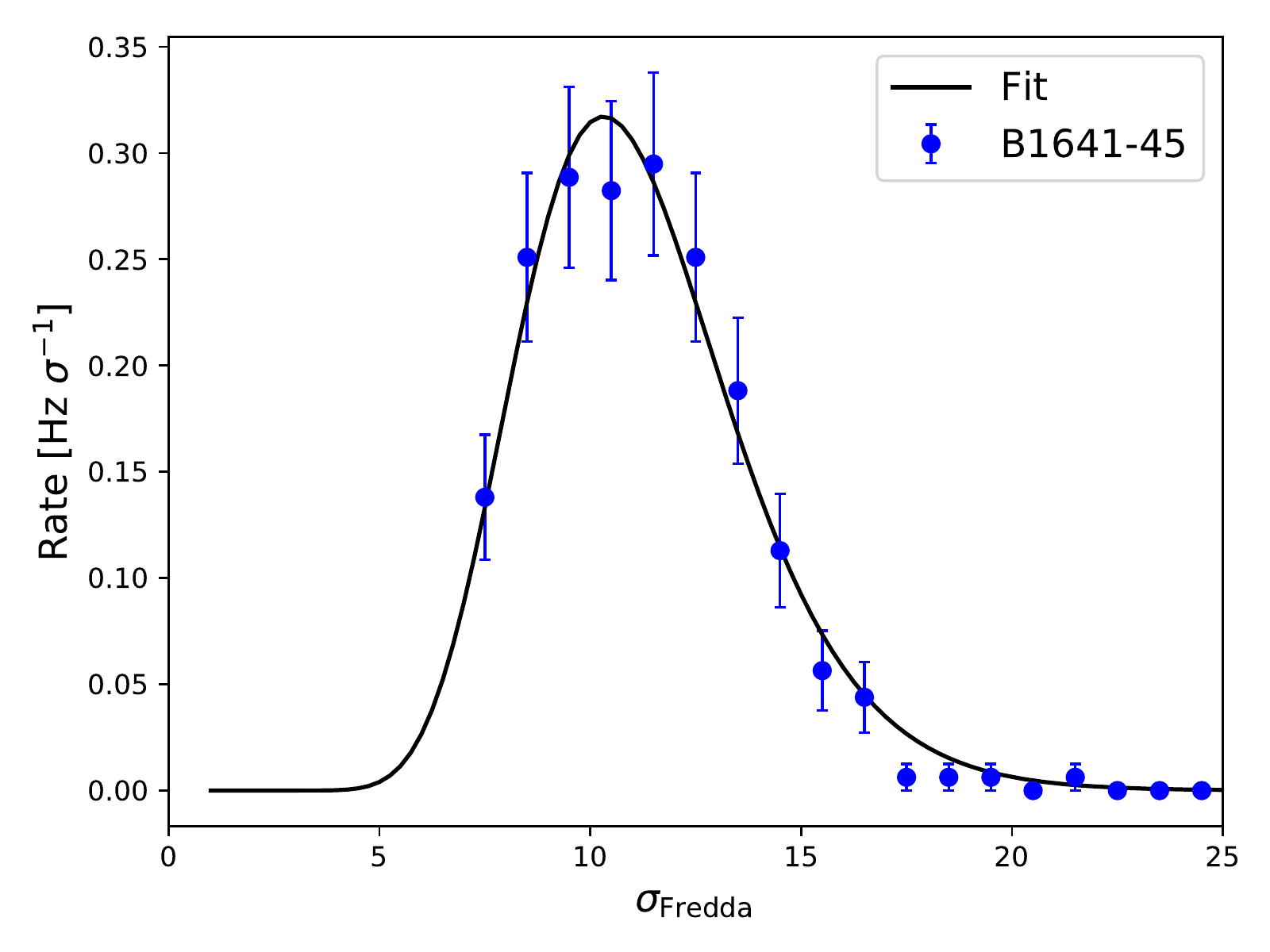}
\caption{Examples of a fits to a pulsar calibration observation, for B0833-45 (left) and B1641-45 (right). Points: histogram of detected pulses from B1641 in a single beam, showing Poisson error bars; line: fit from equation~(\ref{eq:lognormal}). }\label{fig:pulsar_fit}
\end{center}
\end{figure*}

For each beam/antenna/calibration, histograms of $\sigma_{\rm F}$ are normalised to the observation time, producing a rate histogram $R$. The rate is then fitted using a log-normal distribution, which has been found to well-model the pulse amplitude distribution of both B1641-45 \citep{2004MNRAS.353..270C} and B0833-45 \citep{2001ApJ...563L..65C}:
\begin{equation}
R(\sigma_{\rm F}) = \frac{r}{\Delta \sigma_{\rm F} \sqrt{2 \pi}} \exp \left( - \frac{(\log_{10} \sigma_{\rm F} - \mu)^2}{2 (\Delta \sigma_{\rm F})^2} \right). \label{eq:lognormal}
\end{equation}
Here, $r$ measures the total fitted rate of pulses (both above and below the CRAFT detection threshold), $\mu$ is $\log_{10}$ of the characteristic sensitivity, and $\Delta \sigma_{\rm F}$ is the spread. The analysis of \citet{2004MNRAS.353..270C} applies to each time-resolved portion of the pulse profile, at a resolution much smaller than $\Delta t=1.2656$\,ms. However, the reported values of the fitted parameters change only slowly over the pulse profile, so we expect it to be sufficiently applicable.\footnote{The FWHM of B1641-45 is approximately $10$\,ms \citep{2004MNRAS.348.1229J}, and is readily resolved by CRAFT; for B0833-45, it is 1--2\,ms \citep{2001ApJ...549L.101J}, and is marginally resolved. This is why B1641-45 is chosen for absolute calibration against Parkes data in Section~\ref{sec:absolute}.}

The fit procedure uses the Levenberg-Marquardt algorithm, implemented in Python 2.7.14 via the SciPy 1.0.0 function {\tt scipy.optimize.curve\_fit} \citep{scipy}. A bias towards low values of efficiency was found when using Poisson-weighted errors, so all errors were set to unity. Histograms are then multiplied with the bin width in log-space, and divided by the observation time to obtain units of rate per log-interval in $\sigma_{\rm F}$.
An example of a fit is given in Figure~\ref{fig:pulsar_fit}.

When sensitivity is low, only the falling tail of the pulse distribution is observed, and the fit becomes degenerate. To remove this, fits were first performed to estimate $\Delta \sigma_{\rm F}$, and then data was re-fitted leaving only $r$ and $\mu$ free. The correlation of fit errors was typically less than $2\%$, and slightly anti-correlated. This amounts to modelling a constant underlying distribution of pulse strengths for each pulsar, which are then are modified independently by efficiency and sensitivity.

\subsection{Efficiency}
\label{sec:efficiency}

Bursts of RFI, large power spikes, or simply a malfunction in the hardware during the commissioning phase can cause a loss of effective observing time, $T_{\rm eff}$, compared to the total observation time, $T_{\rm obs}$. The observation efficiency $\epsilon$ is thus defined as:
\begin{eqnarray}
\epsilon & \equiv & \frac{T_{\rm eff}}{T_{\rm obs}}
\end{eqnarray}
and can be measured through pulsar calibration observations comparing the fitted pulsar pulse rate, $r$, to the known spin rate $r_0$, taken from PSRCAT\footnote{ http://www.atnf.csiro.au/research/pulsar/psrcat} \citep{2005AJ....129.1993M}:
\begin{eqnarray}
\epsilon & = & \frac{r}{r_0}.
\end{eqnarray}
This assumes that factors leading to a loss of $T_{\rm eff}$ affect both pulsar and FRB searches equally. Note that the fitted value of $r$ reflects the total pulsar rate, i.e.\ it accounts for missed below-threshold pulses.

\begin{figure*}
\begin{center}
\includegraphics[width=0.7\textwidth]{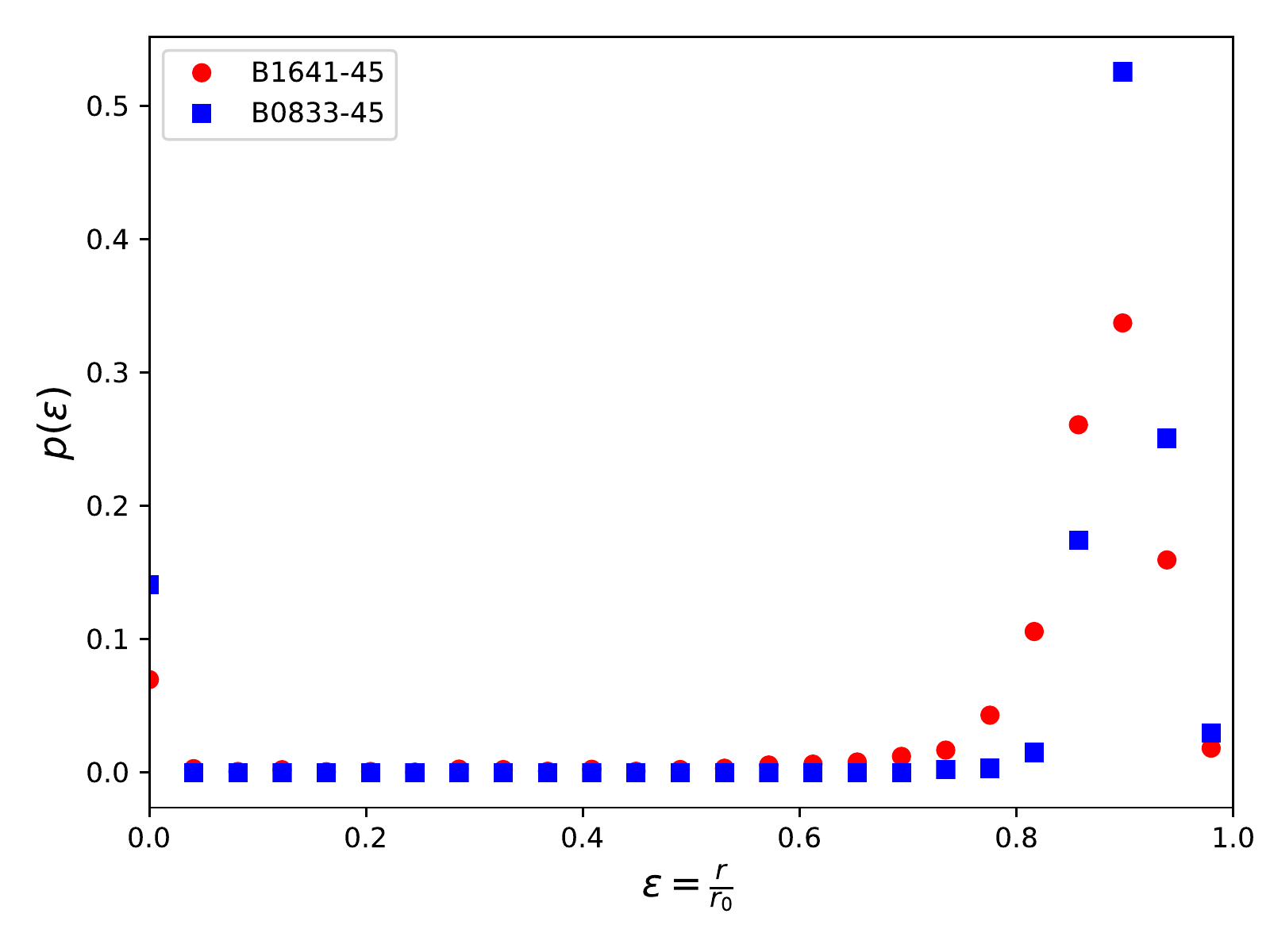}
\caption{Normalised histograms of efficiency for calibration observations of the pulsars B1641-45 and B0833-45, calculated relative to base rates $r_0$ of $2.1975$ and $11.195$\,Hz respectively \citep{2005AJ....129.1993M}. The data is composed of 9,710 independent measurements, and Poissonian errors are too small to be shown on the plot.}\label{fig:rel_rates}
\end{center}
\end{figure*}

Histograms of the fitted rate are given in Figure~\ref{fig:rel_rates} for both B1641-45 and B0833-45. In general, CRAFT efficiencies are in the 80--95\% range. The 6--7\% (85--100 antenna-days equivalent) of data at zero efficiency is partly due to beam $35$ producing unusable data, and partly due to miscellaneous faults during commissioning observations.

The efficiencies measured with B1641-45, $\epsilon_{\rm B1641}$, peak at a similar value (90\%) to those measured with B0833-45, $\epsilon_{\rm B0833}$.
However, they have a slight tail at lower efficiencies. Since the fitted efficiencies are almost uncorrelated with signal strength, it seems this is unlikely to be due to Vela (B0833-45) being much stronger than B1641-45.
While the exact cause of this is unknown, it may be due to different data-taking conditions and antenna performances during calibration runs with each. For instance, most of the data used in this study comes from the second half of 2017, when B0833-45 was more visible during night-time. Furthermore, chirped RFI pulses were present in some ASKAP data, and may have been responsible for some of the variation in efficiency.\footnote{These chirps were due to control system polling of PAF telemetry data such as temperature, voltages etc. The polling system has now been modified to effectively remove these chirps.} An alternative is that Vela is almost 100\% linearly polarised, so that gain offsets between X and Y polarisations would affect the two pulsars differently.

Excluding the values at $0$ (the loss of beam 35 is accounted for in Section~\ref{sec:beams}, while the zero-valued data is treated in Section~\ref{sec:integrated_efficiency}), fits to antenna and beam dependence did not produce statistically significant results, and fitted mean efficiencies varied within $\pm2$\%. No dependency on power fluctuations (see Section~\ref{sec:power_fluctuations}) was observed. Hence, a global mean value of efficiency, $\overline{\epsilon}$, is calculated by averaging results over both pulsars, finding $\overline{\epsilon}=0.87$.

\subsection{Power fluctuations}
\label{sec:power_fluctuations}

\begin{figure*}
\begin{center}
\includegraphics[width=0.7\textwidth]{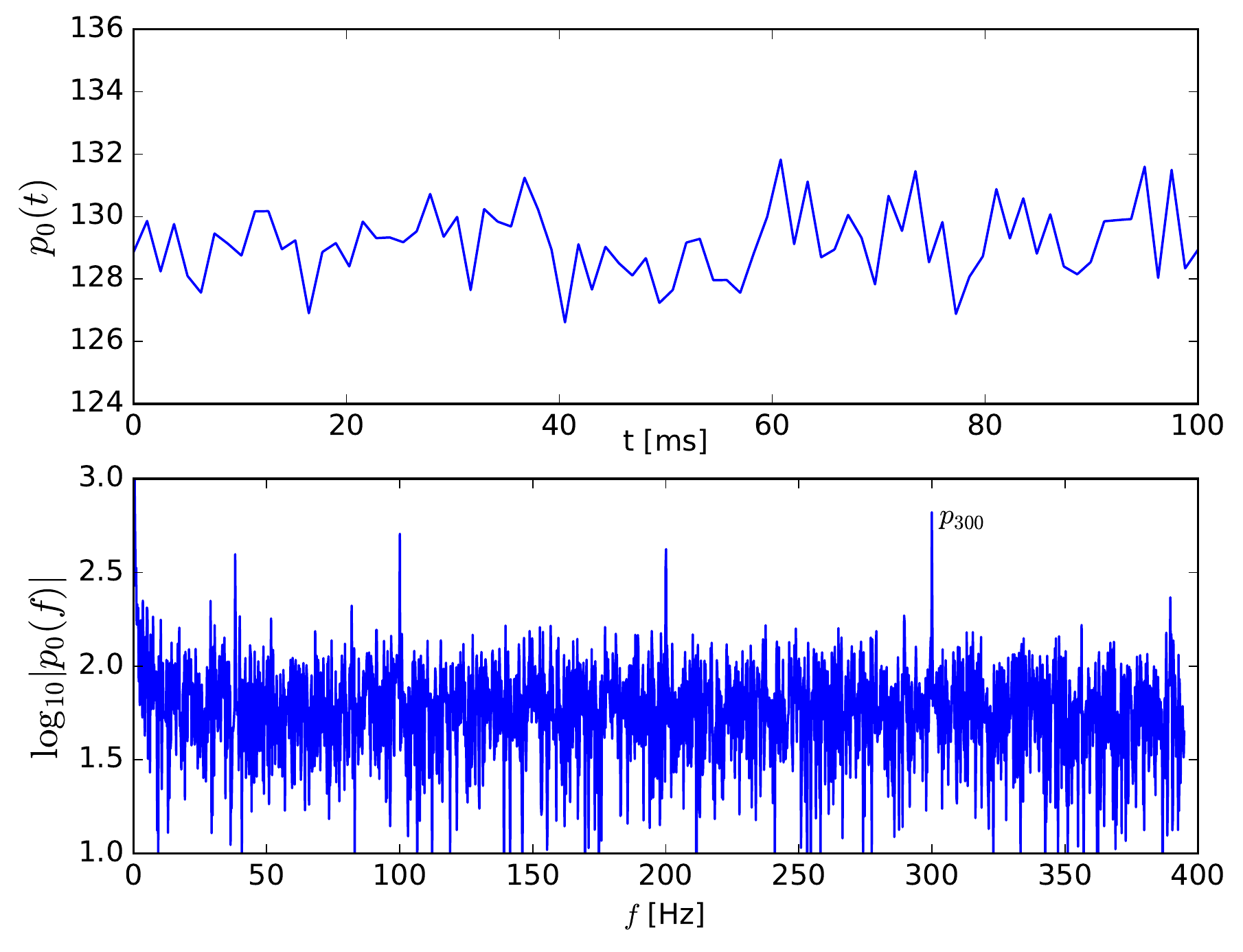}
\caption{Example of power fluctuations in ASKAP commissioning data. Top: example time series of DM0 power $p(t)$, showing the fluctuations over a limited time range. Bottom: Fourier transform magnitude of the time series, taken over 4096 samples. The strength of the peak near 300\,Hz is denoted $p_{300}$. The other peaks are aliased multiples of $300$\,Hz.}\label{fig:300hz_fluc}
\end{center}
\end{figure*}

The only anomaly identified in ASKAP commissioning data was the presence of systematic fluctuations in the digitised CRAFT voltages on timescales of ms and greater. The fluctuations affected all frequency channels uniformly, but grew stronger with increased electrical power requirements, i.e.\ cooling, during daylight hours. It also varied systematically both over the elements of each PAF, and between antennas, according to the power distribution network.

An example of data affected by power fluctuations is given in Figure~\ref{fig:300hz_fluc}, showing the systematic effect when summed over all channels (`DM0' power, $p_0(t)$), and a discrete Fourier transform (DFT) of the DM0 signal, $p_0(f)$.  A strong peak at 300\,Hz is clearly present, with secondary peaks corresponding to aliased multiples of 300\,Hz.

The cause of the problem has now been identified as large voltage fluctuations in the phased array feed power supplies, which has been fixed by adjusting supply voltages. However, two months of CRAFT data (from late June through August 2017) was affected. This coincides with the dearth of FRBs from mid July to August 2017, with no FRBs observed during 280 antenna-days. Equal or longer waiting times occur with a probability of approximately $1$\%; given we sample 19 such waiting times, this observation is not significant, even before accounting for sensitivity reductions, which we do below.

Without modifying the search algorithm, power fluctuations are expected to decrease sensitivity by increasing the system equivalent flux density, and hence the nominal detection threshold. It was expected that most of the original sensitivity could be recovered in offline analysis, either by removing the main Fourier components, or by subtracting the systematic $p_0(t)$ signal from the data (`DM0 subtraction'). Preliminary investigations have found however that both methods produce an equally limited recovery of sensitivity, with a DM0 subtraction method being implemented due to its computational simplicity. This suggests that sensitivity loss is caused at the beamforming stage, by effectively modifying the weights with which PAF elements are summed. The sensitivity loss is therefore deemed unrecoverable --- it is parameterised in the next section.

\subsection{Modelling sensitivity}
\label{sec:modelling_sens}

The relative sensitivity $s_{i,j}$ of CRAFT FRB searches with antenna $i$ and beam $j$ is modelled as:
\begin{eqnarray}
s_{i,j}(p^{\prime}_{300}) & = & a_i b_j P_k n(p^{\prime}_{300}) \label{eq:sens} \\
n(p^{\prime}_{300}) & = & \left[ \left(\frac{p^{\prime}_{300}}{c_1} \right)^2+1 \right]^{-c_2} \label{eq:noise}\\
p^{\prime}_{300} & = & \frac{p_{300}}{p_{\rm med}} 
\end{eqnarray}
where $a_i$ and $b_j$ give the relative sensitivities of antennas $i$ and beam $j$, $P_k$ gives the peak emission of pulsar $k$, $p_{300}^{\prime}$ and $p_{\rm med}$ are respectively the 300\,Hz and median powers illustrated in Figure~\ref{fig:300hz_fluc}, and $c_1$ and $c_2$ are scaling constants. The $b_j$ for each of the two footprints were treated as independent variables, due to their differing sky positions (Figure~\ref{fig:closepack}). The functional form of $n$ was found empirically, with $p^{\prime}_{300}$ being normalised by the median power $p_{\rm med}$ giving the best fit.\footnote{Both $p_{300}$ and $p_{\rm med}$ are calculated using $10$ consecutive DFTs over 1024 samples, with $p_{\rm med}$ being the median value of $|p_0(f)|$ over all non-zero frequencies.}

For the 19 antennas and two footprints used in the search so far, this gave a 93-parameter fit, after constraining the sensitivities of antenna 8 and beam $20$ (one of the central beams in closepack36 configuration) to unity, i.e.\ $a_8=b_{20}=1$. The fit procedure, as in Section~\ref{sec:fitting_method}, also uses the implementation in Python's {\tt scipy.optimize.curve\_fit} function.

Fit errors were estimated with a bootstrapping technique. Pulsar calibration observations were removed one at a time, and used to re-estimate the fit parameters. Since slightly less data is being used in each fit, the resulting variation is too large by a factor of $(n_{\rm cal}/(n_{\rm cal}-1))^{0.5}$, where $n_{\rm cal}$ is the number of calibration observations in which each antenna participated. Some antennas (02, 12, 16, and 30) were only involved in one calibration observation, and hence errors could not be estimated. By definition, however, these antennas contributed very little to observations, and hence this does not greatly affect the average sensitivity.

\subsubsection{Results of the fit}
\label{sec:fit_results}

\begin{figure*}
\begin{center}
\includegraphics[width=0.7\textwidth]{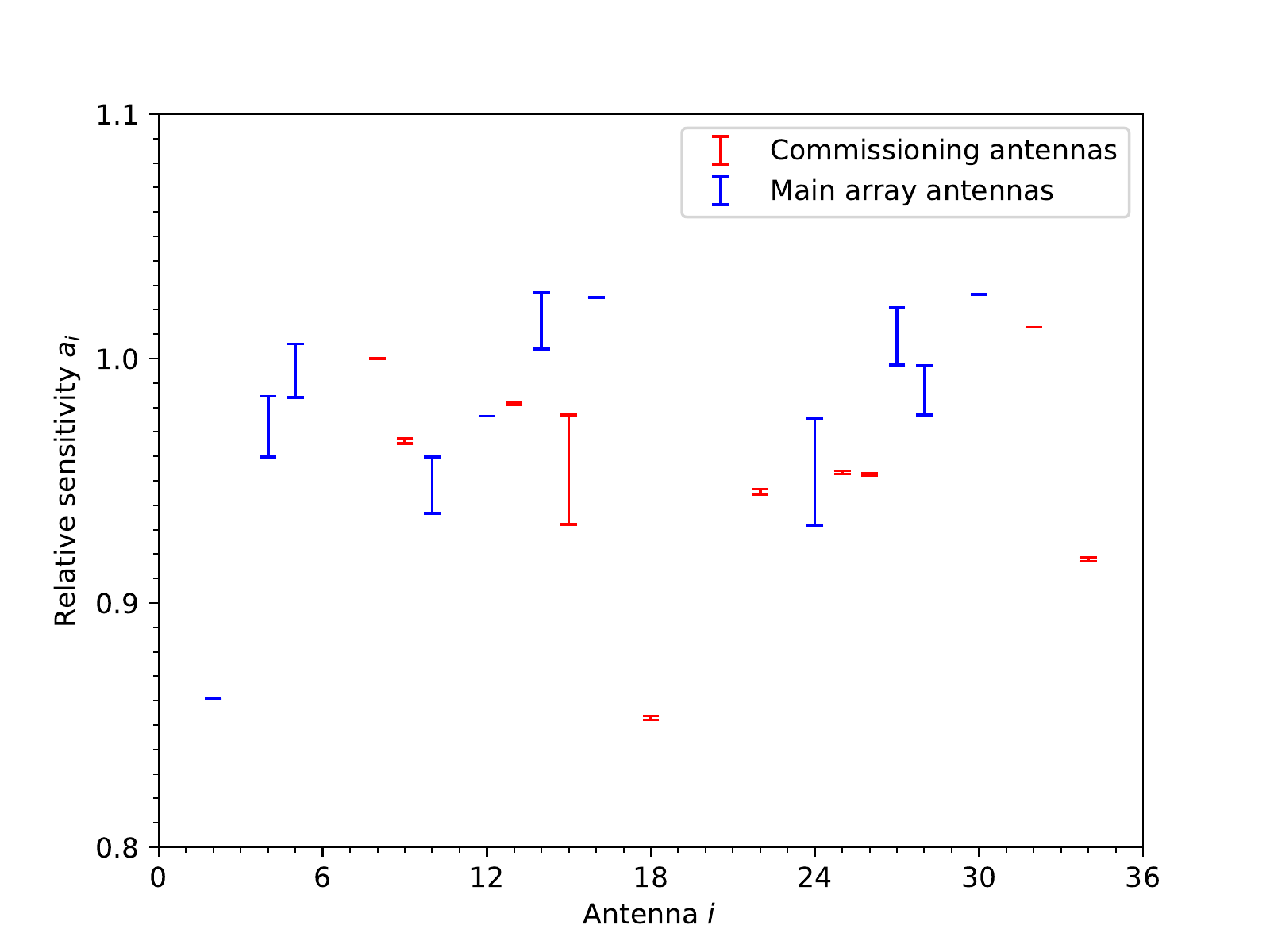}
\caption{Fitted sensitivities $a_i$ for each antenna $i$, relative to antenna 08. No errors could be fitted for antennas 02, 12, 16, 30, and 32, which participated in only a single calibration observation, while main array antennas, and antenna 15, participated in few, leading to larger uncertainty.}\label{fig:afit}
\end{center}
\end{figure*}

\begin{figure*}
\begin{center}
\includegraphics[width=0.7\textwidth]{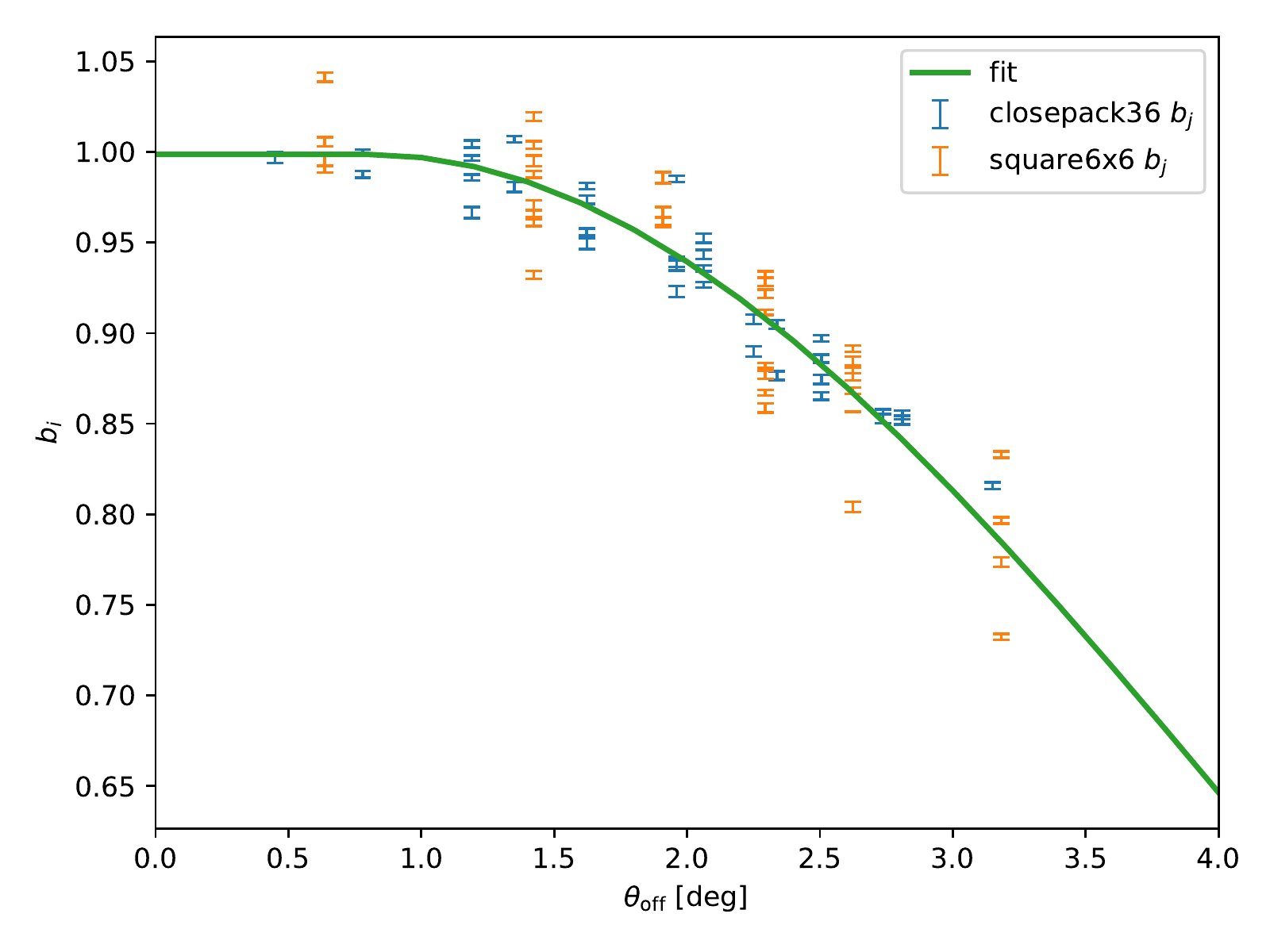}
\caption{Points: relative beam sensitivities $b_i$ from the fit to equation~(\ref{eq:sens}). Line: fit of beam sensitivity as a function of the angular offset $\theta_{\rm off}$ from the antenna optical axis. Main array antennas (blue) have already been commissioned, and are connected to the ASKAP correlator; commissioning antennas are those used during the commissioning period.}\label{fig:bfit}
\end{center}
\end{figure*}

\begin{figure*}
\begin{center}
\includegraphics[width=0.7\textwidth]{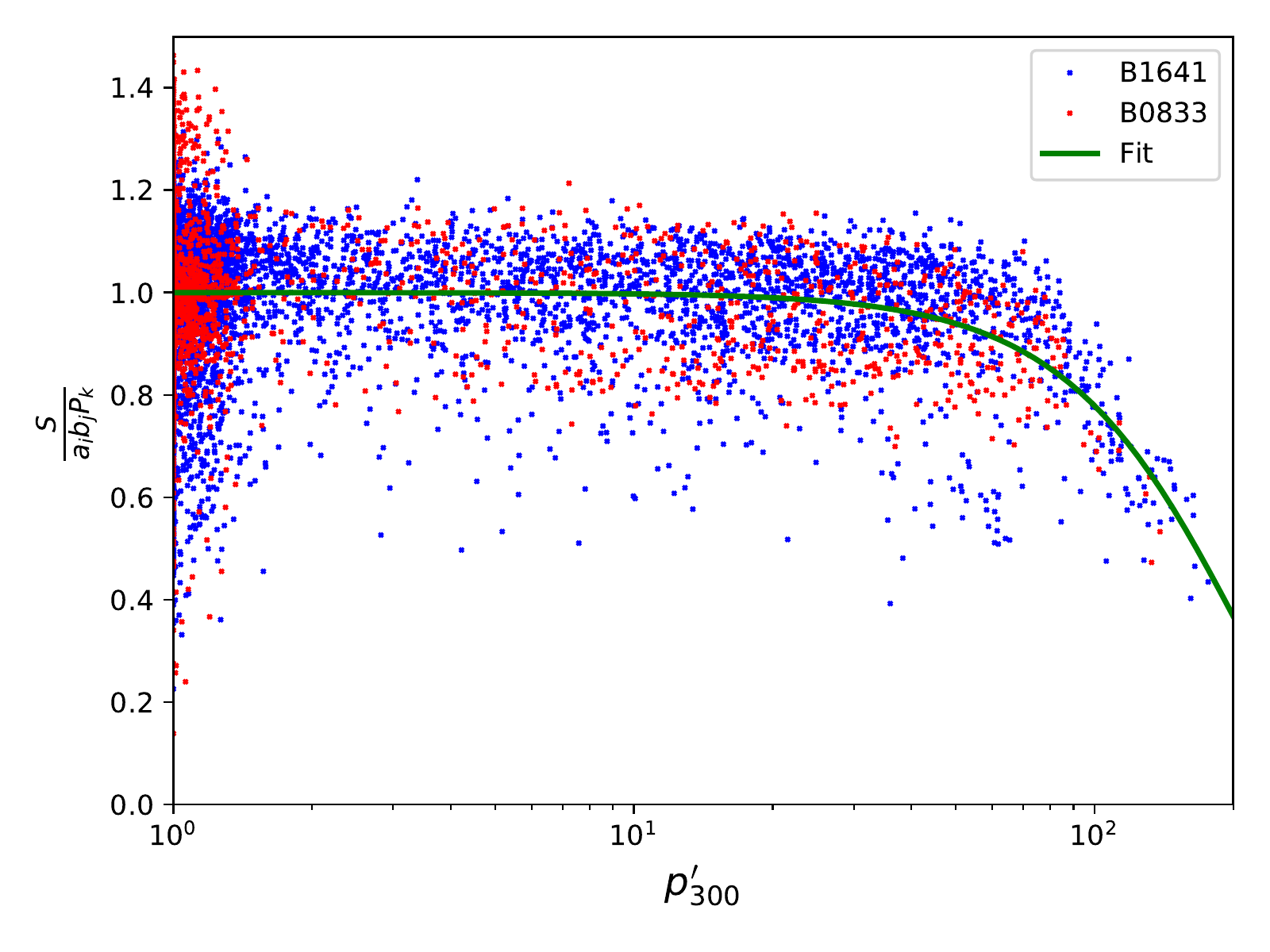}
\caption{The effect of $300$\,Hz noise on CRAFT sensitivity. Points: calibration observations of B1641 (blue) and B0833 (red), after removing the fitted effects of antenna, beam, and pulsar single pulse detection significance (equation~(\ref{eq:sens})), compared to the fitted noise function $n$ (equation~(\ref{eq:noise})).}\label{fig:p300fit}
\end{center}
\end{figure*}

The fitted values of $a_i$, $b_j$, and $n(p_{300}^{\prime})$ are illustrated in Figures \ref{fig:afit}, \ref{fig:bfit} and \ref{fig:p300fit} respectively. The median significances at which single pulses from each pulsar are detected in CRAFT data from antenna 08 beam 20 using FREDDA were found to be $P_{\rm B1641}=12.44 \pm 0.05$ and $P_{\rm B0833}=16.87 \pm 0.13$. The mean antenna sensitivity was found to be 96.7\% that of antenna 08, with rms variation between antennas of $\pm 4.7$\%. This value, calculated on MkII PAFs, compares well with the approximate $\pm 5\%$ variation in $T_{\rm sys}$ and system equivalent flux density (SEFD) found by \citet{ACES005} for a partially overlapping sample of antennas with MkI PAFs.

Beam sensitivities far from the antenna optical axis are expected to fall, due primarily to the finite extent of the PAF. To model this, the values of $b_i$ are fit as a function of beam angular offset, $\theta_{\rm off}$, according to a flattened Gaussian:
\begin{eqnarray}
b(\theta_{\rm off}) & = & \left\{ \begin{array}{ll}
1 & \theta_{\rm off} \le \theta_0 \\
e^{-0.5 \left(\frac{\theta_{\rm off}-\theta_0}{\sigma_{\theta}} \right)^2 }&   \theta_{\rm off} > \theta_0 \\
\end{array}\right.
\end{eqnarray}
The results of this fit are shown in Figure~\ref{fig:bfit}. The fitted beam sensitivities have also been compared to the apodizing function found by \citet{ACES015}, which includes the 2D structure of the PAF. This was found to over-correct the beam sensitivities. A possible cause may be that the measured $b_j$ are functions of both beam shape, peak sensitivity, and mean pointing error. Since beams far from the optical axis are approximately 5\% broader than inner beams, a mis-pointed outer beam will suffer less sensitivity reduction than a mis-pointed inner beam.

From Figure~\ref{fig:p300fit}, the $300$\,Hz power fluctuations cause negligible change in sensitivity, up to the point where $p_{300}^{\prime} \approx 50$, after which the sensitivity falls sharply. The fitted values of $c_1$ and $c_2$ (equation~(\ref{eq:noise})) are $1.22 \cdot 10^4$ and $3700$ respectively, although these values are poorly constrained, highly correlated, and their errors are not well-estimated with the bootstrap method due to the small amount of data with very high values of $p_{300}^{\prime}$.

\subsection{Integrated sensitivity}
\label{sec:integrated_efficiency}

The time-integrated sensitivity of the CRAFT GL50 FRB survey is treated as a function of the sensitivity of the telescopes used (the $a_i$) and power fluctuations, $n(p_{300}^{\prime})$; the fitted values of $b_i$ will be incorporated into the beam model in Section~\ref{sec:beams}.

The relative sensitivity of each antenna  relative sensitivity $s_{i,k}$ for each antenna $i$ and (typically one hour) scan $k$ is calculated via:
\begin{eqnarray}
s_{i,k} & = & \frac{a_i}{35} \sum_{j=0}^{34} n(p_{300,i,j,k}^{\prime})
\end{eqnarray}
where the $p_{300,i,j,k}^{\prime}$ are calculated using the first $10 \times 1024$ samples from each scan. The effect of the virtual loss of beam 35 is accounted for in beamshape estimates (Section~\ref{sec:beams}).

This procedure ignores the fact that power fluctuations do not appear uniformly over the total beam pattern, nor do its effects add linearly between beams. Beams with low sensitivity will be partially compensated for by neighbouring beams, and the effect of an edge beam having its sensitivity reduced will be greater than for an interior beam. However, it has been found that neighbouring beams experience similar amounts of power noise, so these effects will be small, and are ignored.

\begin{figure*}
\begin{center}
\includegraphics[width=0.7\textwidth]{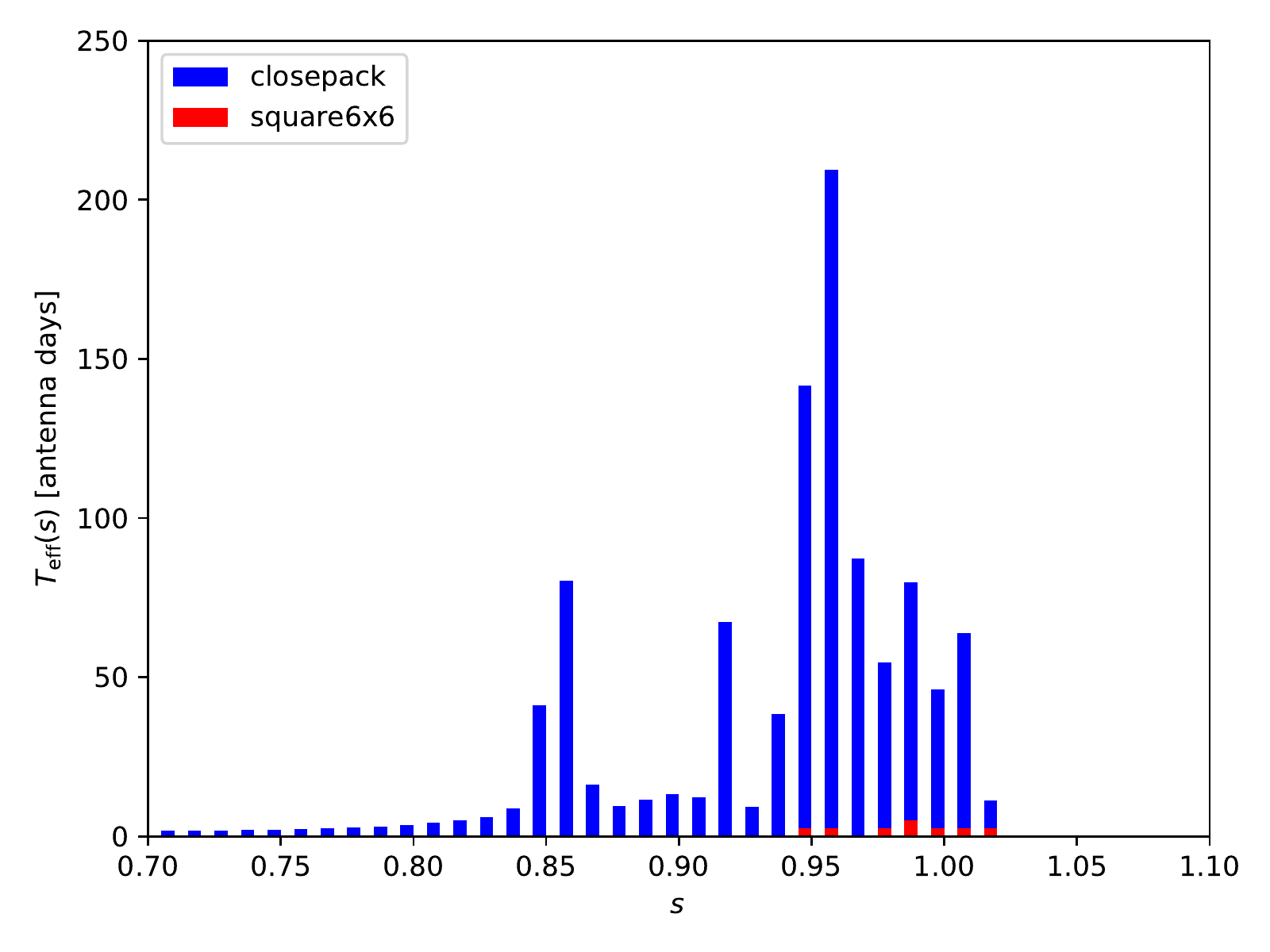}
\caption{Histogram of total observation time at relative sensitivity $s$, divided into contributions from closepack36 (blue) and square6x6 (red) configurations.}\label{fig:sens_hist}
\end{center}
\end{figure*}

To calculate the time-integrated sensitivity, $T_{\rm eff}(s)$, the $s_{i,k}$ are binned with weights equal to the total time-frame of recorded data for each antenna. Due to data losses due to commissioning, the recorded data time was found to be $1274.6$ antenna days, compared with the nominal time of $1326$\,antenna days quoted in \citet{craft_nature}. A further efficiency factor of 87\,\% was also applied to the observing time, as found in Section~\ref{sec:efficiency}. The total effective observation time, adjusted for efficiency losses, was $T_{\rm eff}=1108.9$ antenna days.

Figure~\ref{fig:sens_hist} shows $T_{\rm eff}(s)$. Each `spike' corresponds to an antenna's base sensitivity, with low-sensitivity tails due to the effects of power noise, which were only significant during closepack36 observations. Most antennas, most of the time, suffered negligible noise effects, and hence their total observation time (antenna days) falls within a single bin.

\section{ASKAP Beamshape}
\label{sec:beams}

ASKAP beams are formed from a complex-weighted sum over individual PAF elements. The sum coefficients are determined for each beam, coarse channel, and polarisation independently, using the maximum signal-to-noise algorithm described in \citet{2014PASA...31...41H} and updated by \citet{2016PASA...33...42M}. A total of $36$ beams per antenna can be formed simultaneously, and are used in CRAFT observations.

The resulting beam patterns can deviate significantly from the idealisations shown in Figure~\ref{fig:closepack}. In particular, RFI or malfunctioning PAF elements can result in distorted or mis-pointed beams. Such aberrations will vary with each antenna and new beamforming solution, and accounting for these is crucial in determining the CRAFT sensitivity pattern. Here, the ASKAP beam pattern in both closepack36 and square6x6 configurations are measured using holography scans.

The procedure for measuring ASKAP beamshapes through holography scans is described in \citet{2016PASA...33...42M}, and based on \citet{1977MNRAS.178..539S}. A bright point-source (e.g.\ Virgo A) is placed at the boresight of a reference antenna, and $36$ duplicate beams are formed on that location. All other `measurement' antennas use a standard beam pattern, and are passed through a regular $15\times15$ grid ($0.6^{\circ}$ spacing) of pointing offsets. Each scan is limited by the visibility of the reference source, and each pointing is set to $90$\,s. Orthogonal linear polarisation (X,Y) channels from the reference antenna are correlated with those from each measurement antennas, for all 1\,MHz channels used in the observation.
Time-averaged values of the four correlation products (XX, YY, XY, and YX) are recorded for every antenna, beam, 1\,MHz band, and pointing. In this way, the correlation power at each pointing offset is proportional to the measurement antenna's beam voltage pattern when mirrored through the boresight. This analysis used holography scans 4327 and 4568 for square6x6 and closepack36 configurations respectively, with parameters given in Table~\ref{tbl:holography}.

\begin{table*}
\caption{Holography scan parameters used for beam calibration: the scheduling block (SBID) of the observation, frequency range, reference source, and the antennas used, with the first being the reference antenna for which no beam pattern is calculated.}
\centering
\begin{tabular}{c c c l}
\hline\hline
SBID & Frequencies & Ref & Antennas  \\
SB04327 & 865--1056\,MHz& Virgo A & 2  3  4  6 10 12 14 16 19 27 28 30 \\
SB04568 & 1201--1440\,MHz & Virgo A & 1 2 3 4 5 6 10 12 14 16 17 19 24 27 28 30 \\
\hline\hline
\end{tabular}
\label{tbl:holography}
\end{table*}

\subsection{Measurements of the beam power pattern}
\label{sec:beam_measurements}

The measurement of each beam power pattern proceeds through the following steps.

\begin{enumerate}

\item For each channel, interpolate the real and imaginary components of XX and YY correlation products (Re$(C_{XX})$, Im$(C_{XX})$, Re$(C_{YY})$, Im$(C_{YY})$) from the coarse ($15 \times 15$) measurement grid onto a finer $141 \times 141$ grid. The SciPy v1.0.0 {\tt scipy.interpolate.rectbivariatespline} routine in Python 2.7.14 was used with $5^{\rm th}$ order splines \citep{scipy}; the difference with $3^{\rm rd}$ order splines in both computational time and final result was negligible.

\item Sum the resulting values to produce a total intensity beam power pattern for each channel, i.e.\ $I=$Re$^2(C_{XX})+$Im$^2(C_{XX})+$Re$^2(C_{YY})+$Im$^2(C_{YY})$. Note that since one correlating $X$ or $Y$ factor comes from the reference antenna, the correlation powers must be squared to retrieve the beam power pattern of the measurement antenna.

\item Sum $I$ over all channels, and calculate a first estimate of the beam centre using the peak value of $I$.

\item Scale each channel beam about this centre by its frequency in GHz, to produce maps in units of degrees GHz. Calculate a new average beam over all channels.

\item Correlate each channel beam with the mean beam, and remove those with significantly different shapes.

\item Recalculate a new beam centre and shape using the channels passing the above cuts.

\end{enumerate}

\begin{figure*}
\begin{center}
\includegraphics[width=0.49\textwidth]{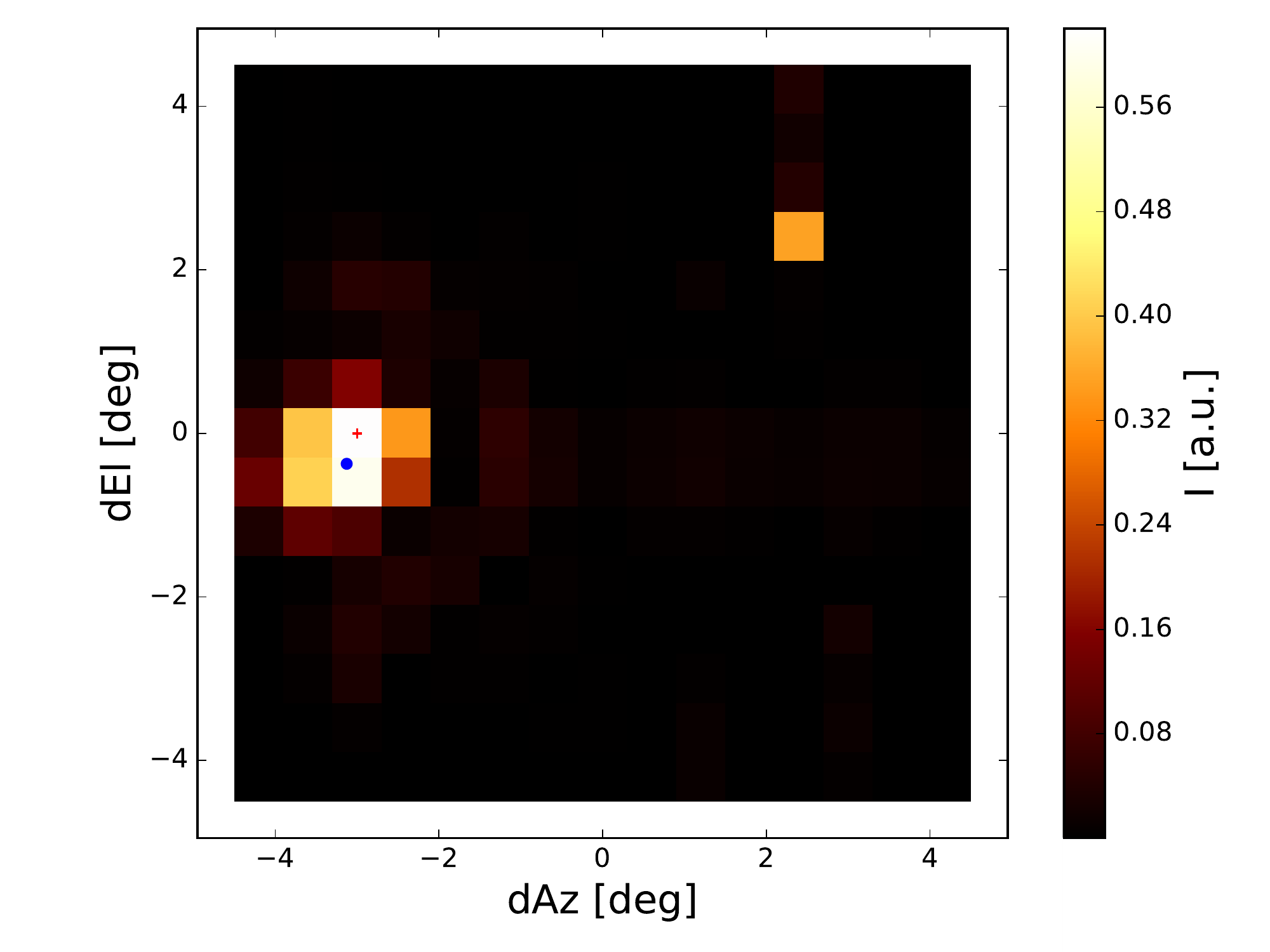} \includegraphics[width=0.49\textwidth]{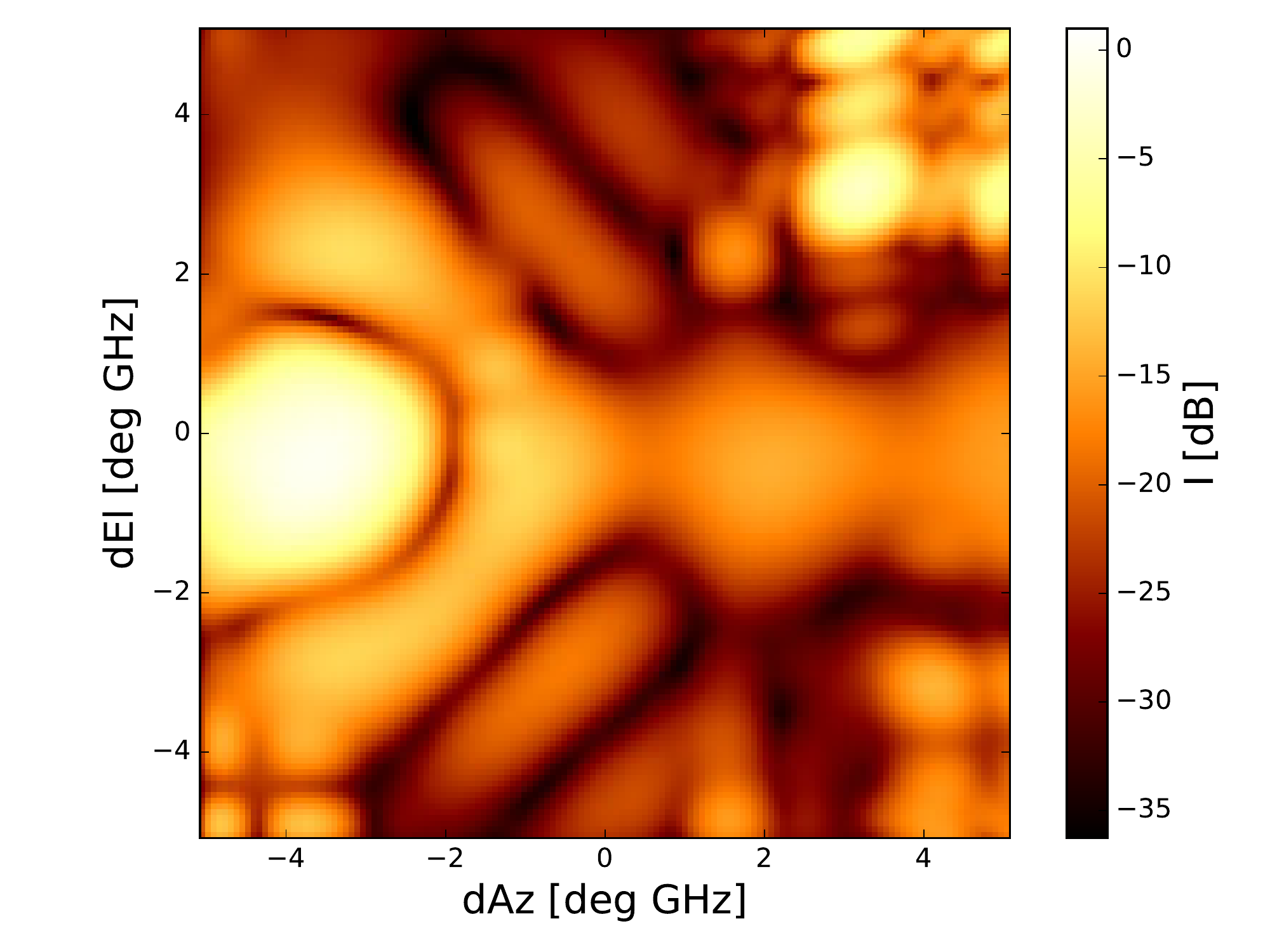}
\includegraphics[width=0.49\textwidth]{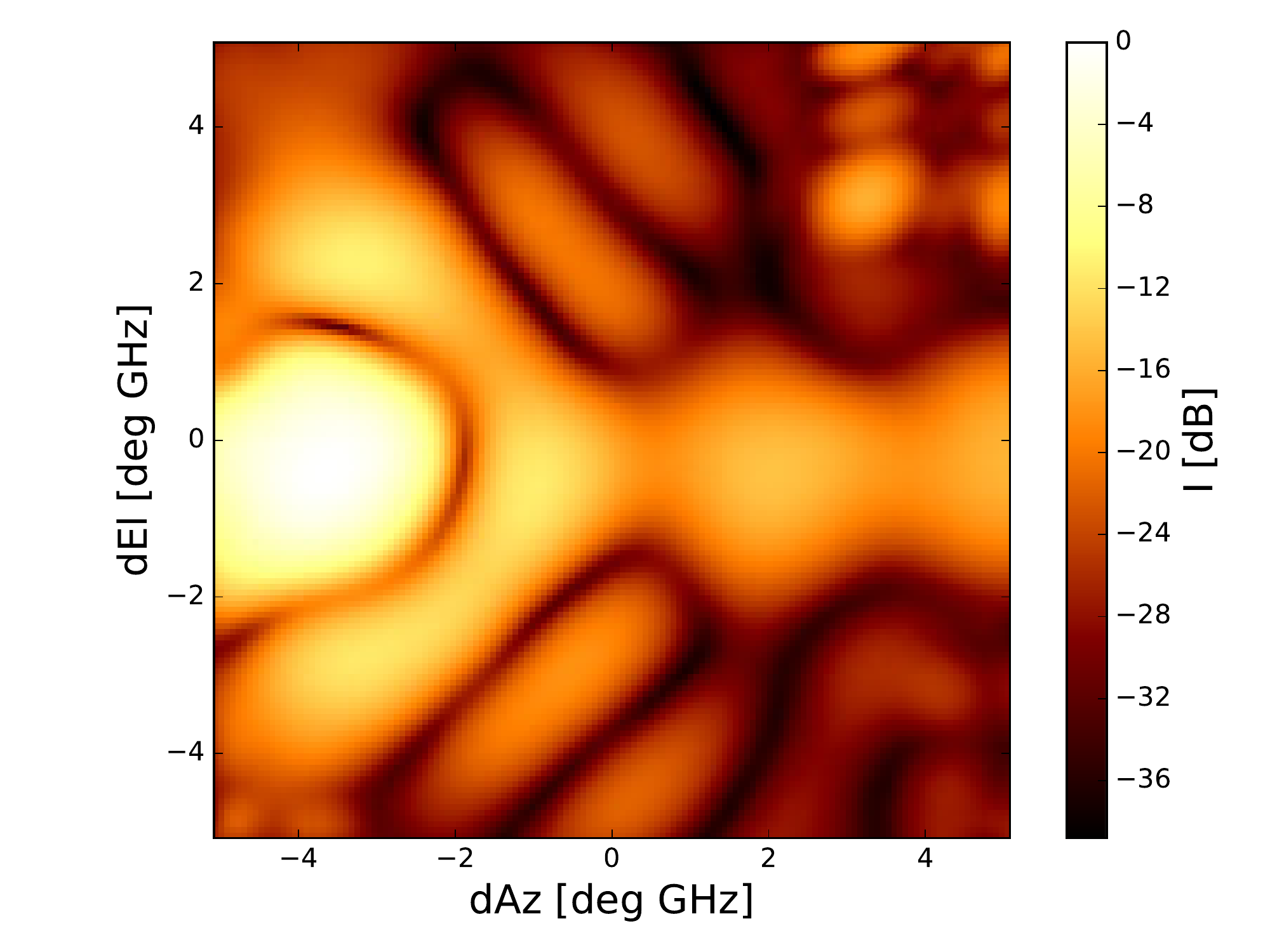}
\includegraphics[width=0.49\textwidth]{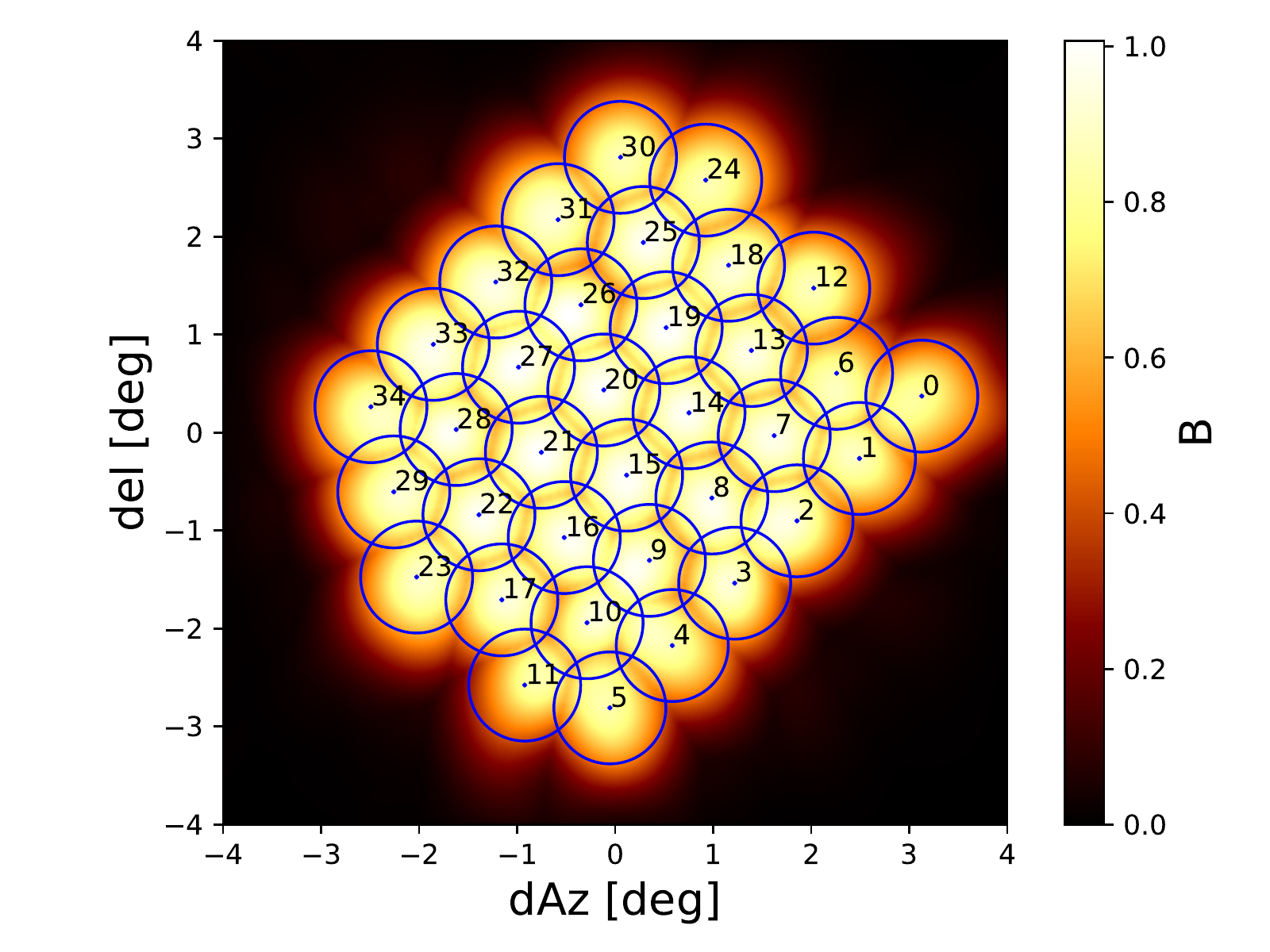}
\caption{Example of the beamshape analysis, for antenna 02, beam 00, in closepack36 configuration. Upper left: raw measurements of total power $I$, averaged over all channels, with blue and red dots showing the expected and first-guess beam centres, respectively. Note the clear presence of RFI near (2.5,2.5) deg. Upper right: interpolated values of $I$ prior to cleaning (`worst case' beam). Lower left: cleaned beamshape (`best case'). Lower right: total closepack36 beamshape in the `best case' scenario, zoomed for clarity. Points indicate the expected beam centres, with circles drawn at the half-power points from an Airy beamshape at 1.296\,GHz. The normalisations are 1: each individual channel has its peak power set to unity prior to averaging; 2,3: the peak value is set to unity, giving a relative beam power pattern; 4: beam $20$ is set to unity, and all other peak beam values are set according to the values of $b_i$ found from the pulsar calibration procedure (Section~\ref{sec:pulsar_cal}). Note that the holography data (top, and lower left, panels) measures the beam position reflected through the origin, which has been corrected-for in the lower right panel.}\label{fig:beamshape}
\end{center}
\end{figure*}

An example of this procedure is given in Figure~\ref{fig:beamshape}. Panel 1 shows the raw measurements in $|C_{XX}|^2+|C_{YY}|^2$ read in at step $1$, averaged over all channels; panel 2 the interpolated values of $I$ calculated at step $3$ prior to cleaning; and panel $3$ shows the cleaned beams at step $6$. The fidelity of each beam can be gauged by the minimum calculated intensity --- typically -40\,dB after step $5$ --- since any effects such as RFI, mis-formed beams, or deviations from the simple width $\propto 1/f$ scaling assumed in Step 4 will act to smear the beamshape over the minimum. The greatest limit to beam fidelity is the use of $I$: repeating the procedure for $XX$ and $YY$ individually resolves much finer structure, but this is not the mode used by CRAFT.

The lower right panel of Figure~\ref{fig:beamshape} compares the measured to expected beamshape. As discussed in Section~\ref{sec:pulsar_cal}, Beam 35 is not plotted, since it's CRAFT data stream was corrupted. Almost all beams are correctly pointed. Outer beams have their peak sensitivity systematically shifted towards centre (consistent with comatic aberration), while for this particular beamset, beam 26 has a notable azimuthal offset.

Since the beam patterns of each footprint are measured only once with a holographic scan, it is ambiguous whether or not any irregularities present are due to mis-formed beams, or peculiarities at the time of observation. In theory, it should be possible to differentiate by modelling the total received power in the scan, where RFI present during the scan will show up as excess power, while mis-formed beams will not. However, the RMS power in each frequency channel is not well constrained, and the integrated values of $XX$ or $YY$ across the grid --- even for `good' channels --- fluctuate significantly about the general trend set by the spectrum of Virgo A, making a definition of `excess' power difficult.

The effect of this ambiguity can be calculated by using both cleaned (step 6) and uncleaned (step 3) beams. These correspond to `best' and `worst' cases, respectively where all irregularities are particular to the holographic scan only, and where they are intrinsic to the beamforming and thus present in CRAFT data.

A further ambiguity is that the holographic scan region extends only $4.2^{\circ}$ in radius, and the beam pattern is typically rotated $45^{\circ}$ to the scan grid, meaning that the sidelobes of corner beams are not measured. This is dealt with here by considering two cases: setting all sidelobes in the unmeasured region to zero, and by estimating the beamshape in the unmeasured region to be equal to the measured beamshape reflected through the beam centre. The former case clearly underestimates the power in the sidelobes, while the latter over-estimates it, since the outer beams tend to be more sensitive in the direction of the boresight.

The total normalisation for each beam is determined using the pulsar calibration observations (see Section~\ref{sec:pulsar_cal}), with outer beams being in general less sensitive than inner beams. As beam 35 is not working, its sensitivity is set to $0$, effectively limiting the CRAFT search to $35$ beams.

Since the candidate search does not combine information from neighbouring beams, the effective sensitivity of CRAFT FRB searches corresponds to the envelope of all remaining $35$ beams. This is calculated for each of the four cases described above. Since sidelobe estimation produces artefacts similar to those in the worst-case scenario, from hereon `sidelobes zero' is synonymous with the `best' case, and `estimated sidelobes' with the `worst' case. An examples of best and worse case beams are given in Figure~\ref{fig:beamshape} (lower left and upper right respectively).

\subsubsection{Effects on CRAFT sensitivity}
\label{sec:sensitivity_effects}

\begin{figure*}
\begin{center}
\includegraphics[width=0.49\textwidth]{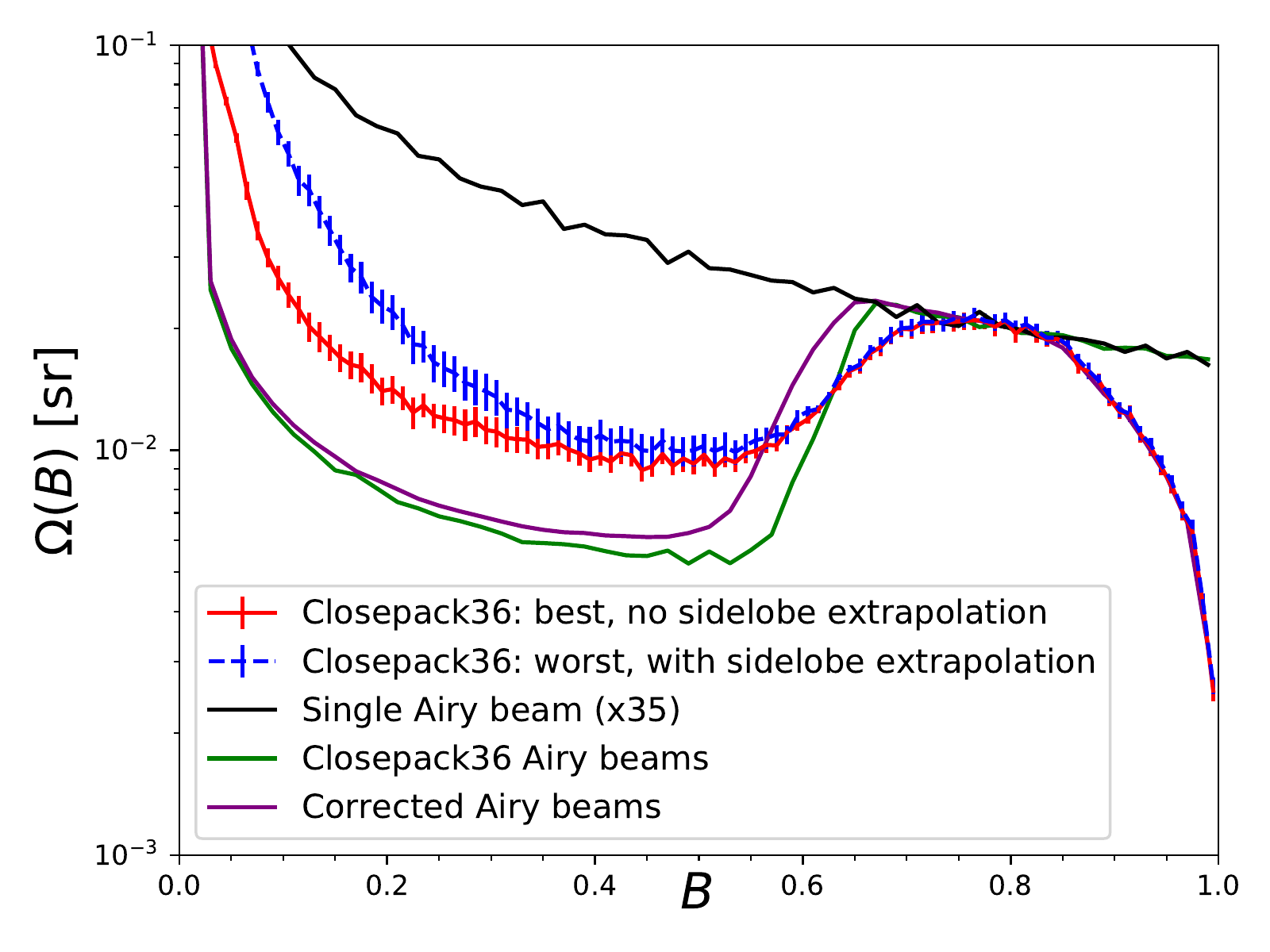} \includegraphics[width=0.49\textwidth]{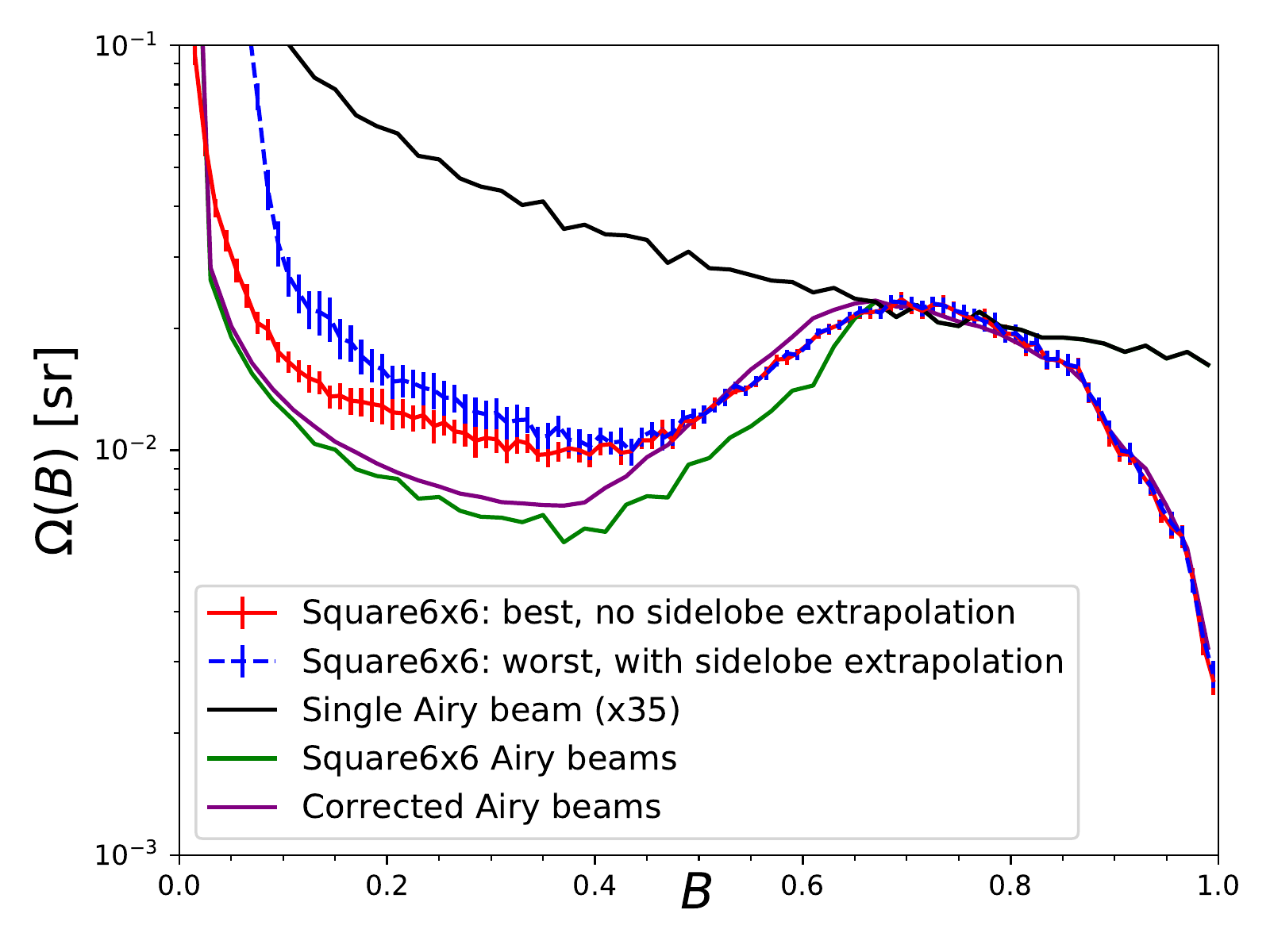} 
\caption{Solid angle $\Omega$ viewed at a given beam sensitivity, $B$, for closepack36 (left) and square6x6 (right) configurations. The black lines show $\Omega$ for 35 (unphysically) independent Airy beams; green shows Airy beams placed at the locations of each ASKAP beam; purple gives the result when these Airy beams are re-normalised to the beam sensitivities found in Section~\ref{sec:fit_results}; and red and blue show the values of $\Omega$ derived from the procedure of Section~\ref{sec:beam_measurements} in both best (red) and worst (blue) case scenarios. The `bumps' in the Airy beam patterns (e.g.\ those in the black line near $B=0.4$) are due to the grid in solid angle used for integrating $\Omega(B)$ as per equation~(\ref{eq:omega_B}), which is identical to that of the ASKAP beam measurements.} \label{fig:beam_sens_hist}
\end{center}
\end{figure*}

The effects of beamshape on sensitivity to FRBs can be characterised by the solid angle of the sky, $\Omega$, viewed with any given sensitivity, $B$, to create a histogram $\Omega(B)$. This can be thought of as inverting the beamshape $B(\Omega)$, although mathematically it is more precisely:
\begin{eqnarray}
\Omega(B) dB  =  \int \, d \Omega \left\{ \begin{array}{ll}
0 & B(\Omega) < B, B(\Omega) \ge B+dB \\
1 & B \le B(\Omega) < B+dB
\end{array} \right. \label{eq:omega_B}
\end{eqnarray}
Observe that characterising a beam by a single value of sensitivity, $B_{\rm eff}$, and solid angle, $\Omega_{\rm eff}$, is equivalent to a `top-hat' beamshape with equal sensitivity of $B_{\rm eff}$ over a solid angle of $\Omega_{\rm eff}$, and zero otherwise. For such a beam, $\Omega(B)$ is:
\begin{eqnarray}
\Omega(B) & = &  \Omega_{\rm eff} \delta(B-B_{\rm eff}) + (4 \pi-\Omega_{\rm eff}) \delta(B)
\end{eqnarray}
where $\delta$ is a Dirac delta function. In general, beams will always view much more of the sky at low sensitivities than high. Both sidelobes, and regions of low primary beam sensitivity, will show up similarly, as large values of $\Omega(B)$ for low $B$.

Figure~\ref{fig:beam_sens_hist} plots $\Omega(B)$ for several cases. The black line shows a simple Airy disk calculated at the mean frequency of 1.296\,GHz for a 12\,m ASKAP antenna, and multiplied by $35$ for comparative purposes. The line increases from right to left, since more of the sky is always viewed at lower sensitivity than high. The same grid is used for calculations as for the beamshapes, which causes both numerical fluctuations, and the value of $\Omega(B)$ at $1$ to be greater than zero (in the analytic case, an infinitesimal amount of the sky is viewed at peak sensitivity).

The green line shows the effect of overlapping beams, by placing each Airy beam at the true pointing positions of each beam for that footprint. The green and black lines are identical up to the point where the beams begin to overlap, below which the solid angle covered by the beams begins to overlap, and the total solid angle scales less than linearly with the number of beams.

The purple line (`Corrected Airy beams') includes the effect of outer beams having reduced sensitivity; the intercept at $B=1$ is much lower, since only a few beams have maximum sensitivity; and the point of overlap is also at lower sensitivity.

Upper and lower bounds on CRAFT sensitivity are shown in red (best case, no sidelobe extrapolation) and blue (worst case, with sidelobe extrapolation). These are calculated by averaging over all antennas in the holography observations --- error bars are the resulting error in the mean, calculated individually for each bin. Differences between antennas dominate the uncertainty in the region of peak sensitivity ($B>0.5$), while systematic effects dominate at low sensitivities ($B < 0.4$). The effect of using a closely packed phased array feed beam footprint is evident in the peakedness of $\Omega(B)$ above $B=0.6$, which is caused by beams obscuring the low-sensitivity regions of neighbouring beams.

Due to holography scans requiring the ASKAP correlator, and CRAFT observations mostly using commissioning antennas which, by definition, are not connected to the correlator, there is no direct way to estimate the beam pattern for all used antennas. Holographic scans are also not performed for every beamforming solution, so that this data set will only be statistically correlated with those used for CRAFT observations. The underlying algorithm for forming PAF beams did remain the same over the course of CRAFT observations. The mean beam- and antenna-dependent factors calculated in Section~\ref{sec:pulsar_cal} are thus assumed to average over these time-dependent factors.

\section{CRAFT SENSITIVITY TO FRBs}
\label{sec:sensitivity}

\subsection{Absolute normalisation}
\label{sec:absolute}

All the above calculations have been of relative sensitivity, specifically by setting that of antenna 8 beam 20 to unity. In order to convert this to an absolute sensitivity, simultaneous observations of B1641 with Parkes (proposal P737) and ASKAP were used to check the absolute sensitivity scale. The observations used the central beam of the Multibeam receiver, with a bandwidth of $256$\,MHz, centred at $1367.5$\,MHz. DSPSR \citep{2011PASA...28....1V} was used to calculate the mean pulse profiles using a total integration time of $T_{\rm int}=180$\,s of Parkes and ASKAP data in an offline analysis.

The flux density scale on Parkes was first calibrated using Hydra A, assuming an emission of 43.1\,Jy at 1400\,MHz and a spectral index of 0.91 \citep{1968ARA&A...6..321S}. The mean flux density at pulse peak, $S_{\rm peak}^{\rm B1641}$, from B1641 was found to be $15.2 \pm 0.1$\,Jy. This allowed the RMS sensitivity of ASKAP, $S_{\rm rms}$, to be calculated using the numerical profile values at the peak and off-pulse, $S_{\rm peak}^{\rm num}$ and $S_{\rm off}^{\rm num}$, giving:
\begin{eqnarray}
S_{\rm rms} & = & S_{\rm peak}^{\rm B1641} \frac{S_{\rm off}^{\rm num}}{S_{\rm peak}^{\rm num}}. \label{eq:Srms}
\end{eqnarray}
Inverting the radiometer equation for the observation bandwidth $\Delta \nu$ and integration time in each profile bin, $T_{\rm int}/N_{\rm bin}$, the system equivalent flux density (SEFD) for CRAFT (total intensity) data can be calculated as:
\begin{eqnarray}
{\rm SEFD} & = & S_{\rm rms} \sqrt{2 \Delta \nu \, T_{\rm obs}}. \label{eq:SEFD}
\end{eqnarray}

\begin{figure*}
\begin{center}
\includegraphics[width=0.49\textwidth]{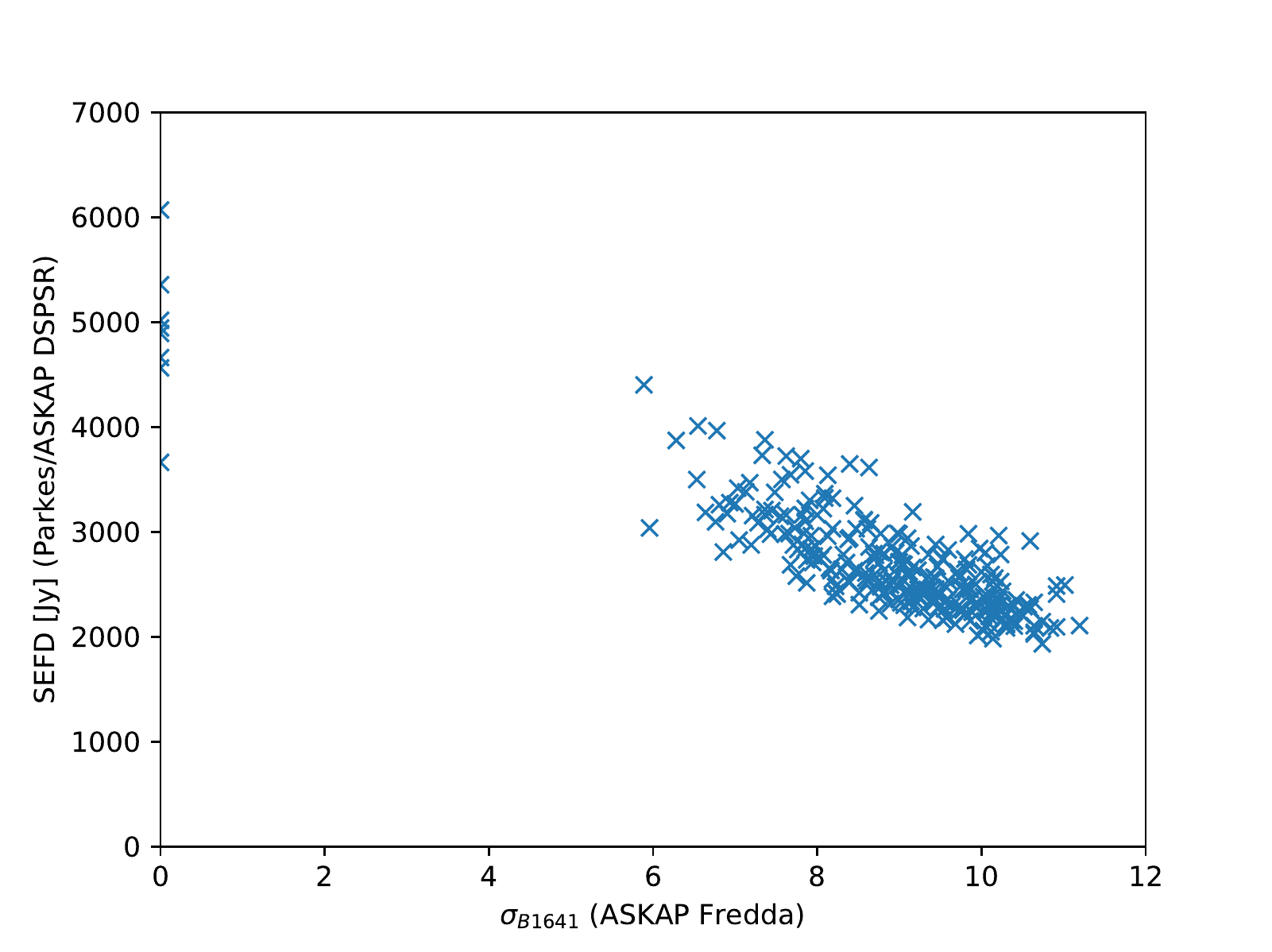} \includegraphics[width=0.49\textwidth]{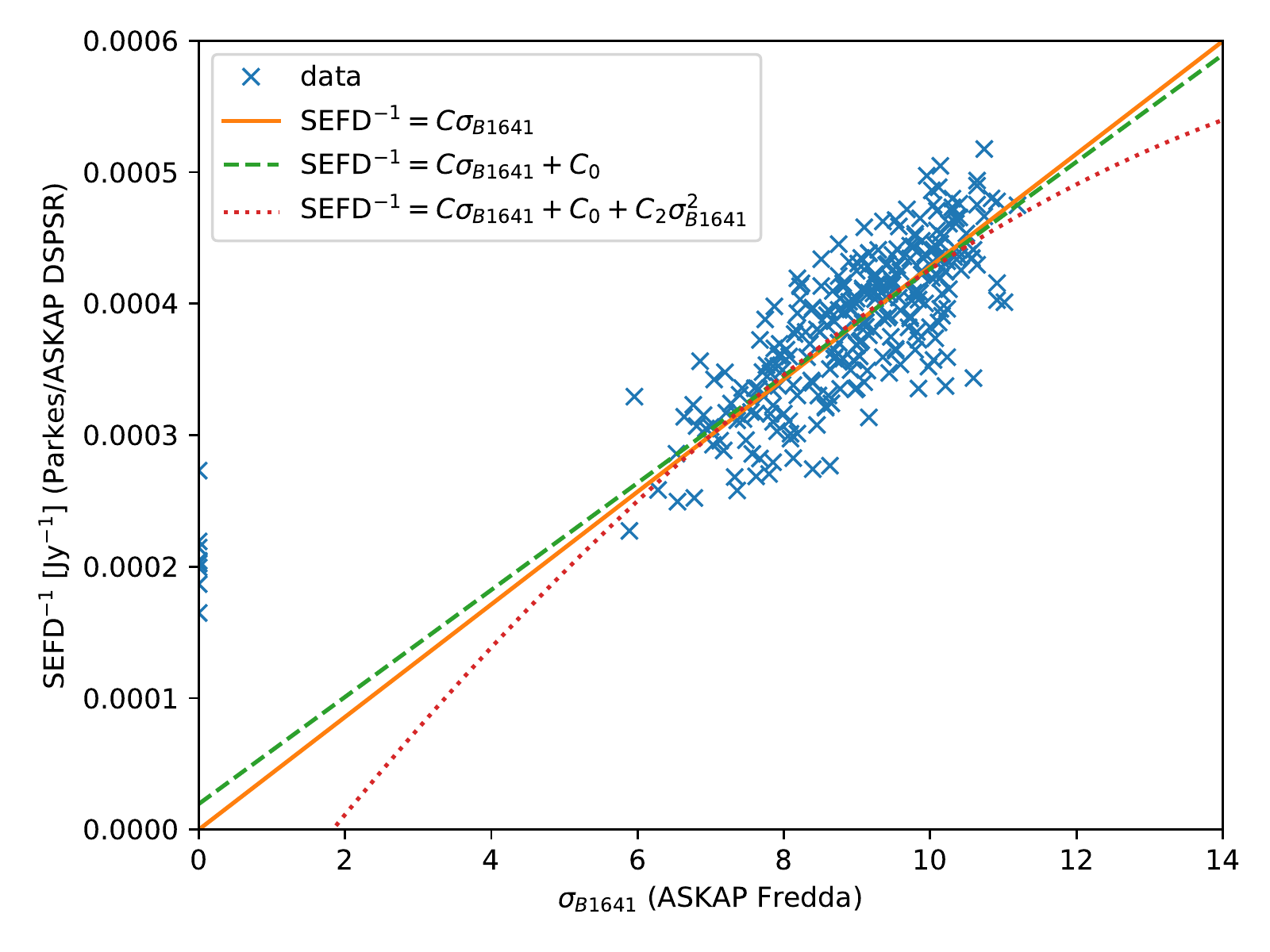} 
\caption{Left: measured sensitivity of ASKAP CRAFT observations, showing system equivalent flux density (SEFD) derived through Parkes--ASKAP observations plotted as a function of CRAFT FREDDA mean pulse height (equation~(\ref{eq:lognormal})) for each antenna--beam. Right: fits of inverse SEFD as a function of mean pulse height, for different functional forms. The data at $\sigma_{\rm B1641}=0$ are from beam 35 and have been excluded from the fit.}\label{fig:absolute_calibration}
\end{center}
\end{figure*}

The values of the SEFD calculated from equation~(\ref{eq:SEFD}) for each antenna/beam are shown in Figure~\ref{fig:absolute_calibration} (left), plotted against the fitted mean sensitivity output by FREDDA from ASKAP observations (equation~(\ref{eq:lognormal})). The lowest SEFDs (obtained for the central beams) are approximately equal to both preliminary SEFD measurements for ASKAP MkII PAFs of $2000$\,Jy \citep{7297174} and optimal values of MkI PAFs \citep{2016PASA...33...42M}. Beam 35, which detected no pulses when passed through FREDDA (hence $\sigma_{\rm B1641}=0$), was found to have a high, but not infinite, SEFD ranging from 3700--6100\,Jy when analysed with DSPSR.

The SEFD of the data is expected to be inversely proportional to the sensitivity of FREDDA, i.e.:
\begin{eqnarray}
{\rm SEFD}^{-1} & = & C \sigma_{\rm B1641} \label{eq:sefd_fit}
\end{eqnarray}
for some constant $C$. This fit is shown in Figure~\ref{fig:absolute_calibration} (right), where beam 35 data has been excluded. Non-linear effects were tested-for by also fitting $1^{\rm st}$ and $2^{\rm nd}$ order polynomials to this range. These did not significantly improve the fits, and in particular, the concave shape of the $2^{\rm nd}$ order fit showed no evidence for an underestimation of the true SEFD when $\sigma_{\rm B1641}$ obtained from FREDDA is low (which would have explained the lack of detections in beam 35).

The cause of the variation in SEFD for a given $\sigma_{\rm B1641}$ is unknown. However, the absolute calibration was performed at a time of high power noise, and it is entirely possible that this induced different responses from DSPSR and FREDDA. A systematic evaluation of pulse search software on common data would contribute greatly to our understanding of this difference.

The constant of proportionality $C$ thus found is $(4.28 \pm 0.02) \cdot 10^{-5}$\,Jy$^{-1}$, i.e.\ $\mbox{SEFD}=2.34 \cdot 10^4 \sigma_{\rm B1641}^{-1}$\,Jy. Given the value of $12.44 \pm 0.05$ fitted for $P_{\rm B1641}$ (the mean value of $\sigma_{\rm B1641}$; see Section~\ref{sec:fit_results}), this means that the normalised SEFD $S_0$ in Section~\ref{sec:pulsar_cal} at beam sensitivity $B=1$ used in Section~\ref{sec:beams} corresponds to an SEFD of $1878 \pm 12$\,Jy. The antenna-averaged value is $1942\pm 12$\,Jy, which agrees well with the nominal value of $2000$\,Jy. It should be noted however that the uncertainty reflects only the random uncertainty of the fitted means --- the variation about the means present in Figure~\ref{fig:absolute_calibration} remains an unexplained systematic effect.

\subsection{Detection threshold}
\label{sec:FRB_sensitivity}

The nominal fluence detection threshold $F_0$ to a perfectly de-dispersed FRB with duration contained entirely within the integration time $t_{\rm int}=1.2656$\,ms is:
\begin{eqnarray}
F_{0} & = & \frac{\sigma_{\rm th} \mbox{SEFD}}{\sqrt{2 t_{\rm int} \Delta \nu}}. \label{eq:S0}
\end{eqnarray}
For the CRAFT bandwidth of $\Delta \nu=336$\,MHz, time resolution of $t_{\rm int}$ 1.2656\,ms, SEFD of $1890\pm13$\,Jy, and detection threshold $\sigma_{\rm th}=9.5$, this corresponds to $F_0=24.6\pm0.2$\,Jy\,ms ($25.5\pm0.2$\,Jy\,ms antenna-average, i.e.\ consistent with the nominal value of $26$ quoted in \citet{craft_nature}). The actual threshold will differ from this value for any real FRB and search method, as discussed in Section~\ref{sec:discussion}.

\begin{figure*}
\begin{center}
\includegraphics[width=0.49\textwidth]{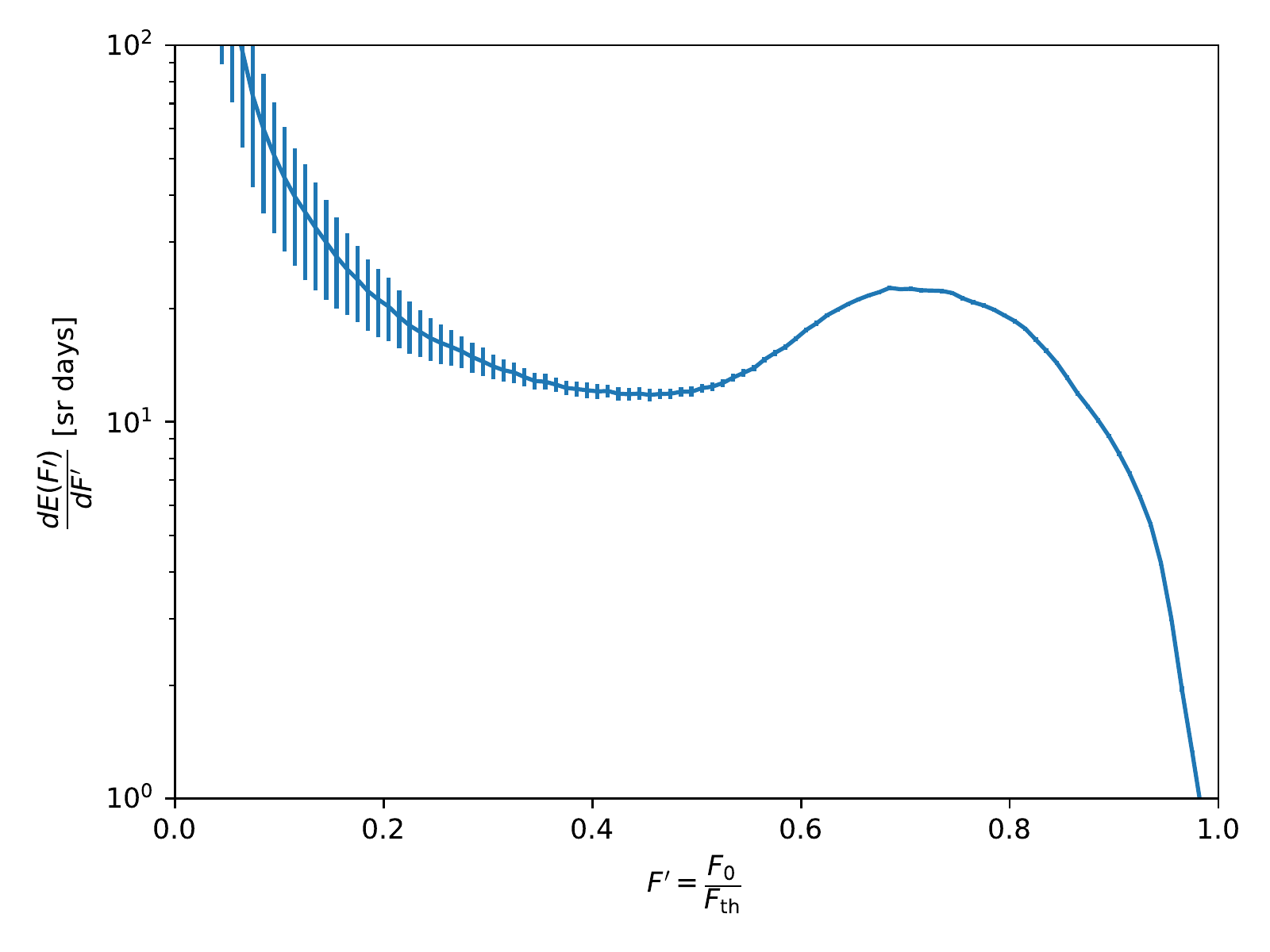} 
\caption{Exposure $E$ of the CRAFT high Galactic latitude survey in terms of relative sensitivity $F^\prime=F_0/F_{\rm th}$, defined such that the integral over $F$ comes to the corrected observation time $T^{\prime}=1208$ days. The mean value is calculated using the average of best case and worst case beam estimated (Figure~\ref{fig:beamshape}), while errors are calculated from the systematic difference between the mean and these cases, added in quadrature to the random uncertainty in the mean.} \label{fig:craft_exposure}
\end{center}
\end{figure*}

The total CRAFT exposure $E$ as a function of fluence threshold $F_{\rm th}$ can be accounted for by integrating the beam-dependence $\Omega(B)$ from equation~(\ref{eq:omega_B}) (displayed in Figure~\ref{fig:beam_sens_hist}) with the time- and antenna-dependence $T(S^{\prime})$ given in Figure~\ref{fig:sens_hist}, and applying the normalisation $F_0=24.8 \pm 0.2$\,Jy\,ms. Summing this over both closepack36 `cp' and square6x6 `sqr' periods produces the survey exposure $E$ as a function of fluence threshold $F_{\rm th}$:
\begin{eqnarray}
E(F_{\rm th}) & = & \sum_{i={\rm cp,sqr}} \int dB \, \Omega_{i}(B) \, T_i(S^{\prime}) \nonumber \\
S^{\prime}& = & B \frac{F_0}{F_{\rm th}}.
\end{eqnarray}
This is given in Figure~\ref{fig:craft_exposure}, in terms of $\frac{dE}{dF}$, and relative fluence sensitivity $F^\prime = \frac{F_0}{F_{\rm th}}$. Note that the contribution from the square6x6 configuration is small, since the majority of observations were made in closepack36 configuration --- Figure~\ref{fig:beam_sens_hist} gives a much better comparison of the relative sensitivities of the two configurations.

\subsection{Effective survey parameters}
\label{sec:effective}

For an FRB population with a power-law spectrum of fluences, such that the detection rate $R$ has the form:
\begin{eqnarray}
R(F_{\rm th},\alpha)  & = & k \left(\frac{F_{\rm th}}{F_0}\right)^{\alpha} [{\rm sky}^{-1} {\rm day^{-1}}] \label{eq:frb_rate}
\end{eqnarray}
for some constant $k$, fluence threshold $F_{\rm th}$ relative to some value $F_0$, and spectral index $\alpha$, the number $N$ of detected FRBs in a survey will be:
\begin{eqnarray}
N(\alpha) & = & \int dF_{\rm th} \, E(F_{\rm th}) \, R(F_{\rm th},\alpha). \label{eq:nalpha_e_r}
\end{eqnarray}
In cases where the only modelled sensitivity dependence is the beamshape, equation~(\ref{eq:nalpha_e_r}) reduces to:
\begin{eqnarray}
N(\alpha) & = & T_{\rm eff}  \int dF_{\rm th} \, \Omega(B=\frac{F_0}{F_{\rm th}}) \, R(F_{\rm th},\alpha) \label{eq:nalpha_t_omega_r} \\
& = & T_{\rm eff}  \int d\Omega \, B(\Omega) \, R(F_{\rm th}=F_0/B,\alpha) \label{eq:nalpha_t_b_r}
\end{eqnarray}
for effective observation time $T_{\rm eff}$, and $B$ and $\Omega(B)$ defined according to equation~(\ref{eq:omega_B}).

Comparisons between different FRB surveys however tend to characterise each in terms of a single fluence threshold $F_{\rm eff}(\alpha)$ and exposure $E_{\rm eff}(\alpha)$, which are both functions of $\alpha$ as discussed by \citet{2018MNRAS.474.1900M}. Defining these to keep $N$ constant:
\begin{eqnarray}
N(\alpha) & = & E_{\rm eff}(\alpha) \, R(F_{\rm eff},\alpha) \label{eq:eff}
\end{eqnarray}
and separating exposure into effective observation time $T_{\rm eff}$ (here antenna days) and solid angle $\Omega_{\rm eff}(\alpha)$:
\begin{eqnarray}
E_{\rm eff}(\alpha) & = & T_{\rm eff} \, \Omega_{\rm eff}(\alpha)
\end{eqnarray}
produces an ambiguity between effective sensitive area $\Omega_{\rm eff}$ and effective threshold $F_{\rm eff}$.

There are three natural constraints to remove this ambiguity. The simplest is to set the threshold $F_{\rm eff}$ equal to the nominal threshold $F_0$ at beam centre, and modify the exposure accordingly, i.e.:
\begin{eqnarray}
\Omega_{\rm eff}(\alpha) & = & T_{\rm eff}^{-1}\int dF_{\rm th} \, E(F_{\rm th}) \frac{R(F_{\rm th},\alpha)}{R(F_0,\alpha)} \nonumber \\
& = & T_{\rm eff}^{-1} \int dF_{\rm th} \, E(F_{\rm th}) \left( \frac{F_{\rm th}}{F_0} \right)^{\alpha}. \label{eq:direct_e_eff}
\end{eqnarray}
This figure can be misleading, however, because $F_0$ is not characteristic of the detected FRB fluences.

This effect can be accounted for by defining $F_{\rm eff}(\alpha)$ as being equal to the mean threshold $\bar{F}_{\rm th}(\alpha)$:
\begin{equation}
\bar{F}_{\rm th}(\alpha) = \frac{1}{N(\alpha)} \int dF_{\rm th} \, F_{\rm th} \, E(F_{\rm th}) \, R(F_{\rm th},\alpha) \label{eq:fbar}
\end{equation}
or to the mean true fluence $\bar{F}_{\rm true}$ of detected FRBs:
\begin{eqnarray}
\bar{F}_{\rm true}(\alpha) & = & \frac{1}{N(\alpha)} \int dF_{\rm th} E(F_{\rm th})
\int_{F_{\rm th}}^{\inf} dF^\prime F^\prime \frac{d R(F^\prime,\alpha)}{dF^\prime} \nonumber \\
& = & \frac{1}{N(\alpha)} \int dF_{\rm th} E(F_{\rm th}) \frac{\alpha}{\alpha+1} F_{\rm th} R(F^\prime,\alpha) \nonumber \\
& = & \frac{\alpha}{\alpha+1} \bar{F}_{\rm th}(\alpha).
\end{eqnarray}
The difference between these two measures only becomes important when models deviating from a pure power law are being fitted, in which case experimental sensitivity should not be reduced to a single effective threshold.

Here, we choose $F_{\rm eff} \equiv \bar{F}_{\rm th}$. This then defines $\Omega_{\rm eff}(\alpha)$ through equation~(\ref{eq:eff}), or by replacing $F_0$ by $\bar{F}_{\rm th}$ in equation~(\ref{eq:direct_e_eff}).

It is important to note that equation~(\ref{eq:fbar}) becomes ill-defined as $\alpha$ approaches unity, which can be seen through the dependence of the integrand on $\alpha$:
\begin{equation}
\bar{F}_{\rm th}(\alpha) \propto \int dF_{\rm th} \, F_{\rm th}^{\alpha+1} \, E(F_{\rm th}). \label{eq:fbar_problem}
\end{equation}
Even as $F_{\rm th}$ approaches zero, its contribution to the integrand remains constant when $\alpha=-1$, and the integral in equation~(\ref{eq:fbar_problem}) evaluates to $4 \pi T_{\rm eff}$ (i.e.\ the total exposure at all sensitivities). This can be understood as a very small number of expected detections at very low sensitivity (e.g.\ in a beam's far sidelobes) contributing correspondingly large values of true fluence threshold $F_{\rm th}$. As $\alpha$ approaches $-1$, the result for $\bar{F}_{\rm th}$ becomes highly dependent on numerical details, e.g.\ histogram bin width in the exposure function $E$, and the distance out to which $\Omega(B)$ is evaluated. This is not the case for the choice of $F_{\rm eff} \equiv F_0$, where $\Omega_{\rm eff}$ (equation~(\ref{eq:direct_e_eff})) remains defined for $\alpha < 0$. However, this should be viewed as an advantage, serving as a reminder that as $\alpha$ approaches $-1$, the true fluences of detected FRBs become poorly correlated with any choice of effective threshold $F_{\rm eff}$, so that the ease of calculation for the choice $F_{\rm eff} \equiv F_0$ merely provides a false sense of security.

\subsubsection{Effective survey parameters: results}

\begin{figure*}
\begin{center}
\includegraphics[width=0.49\textwidth]{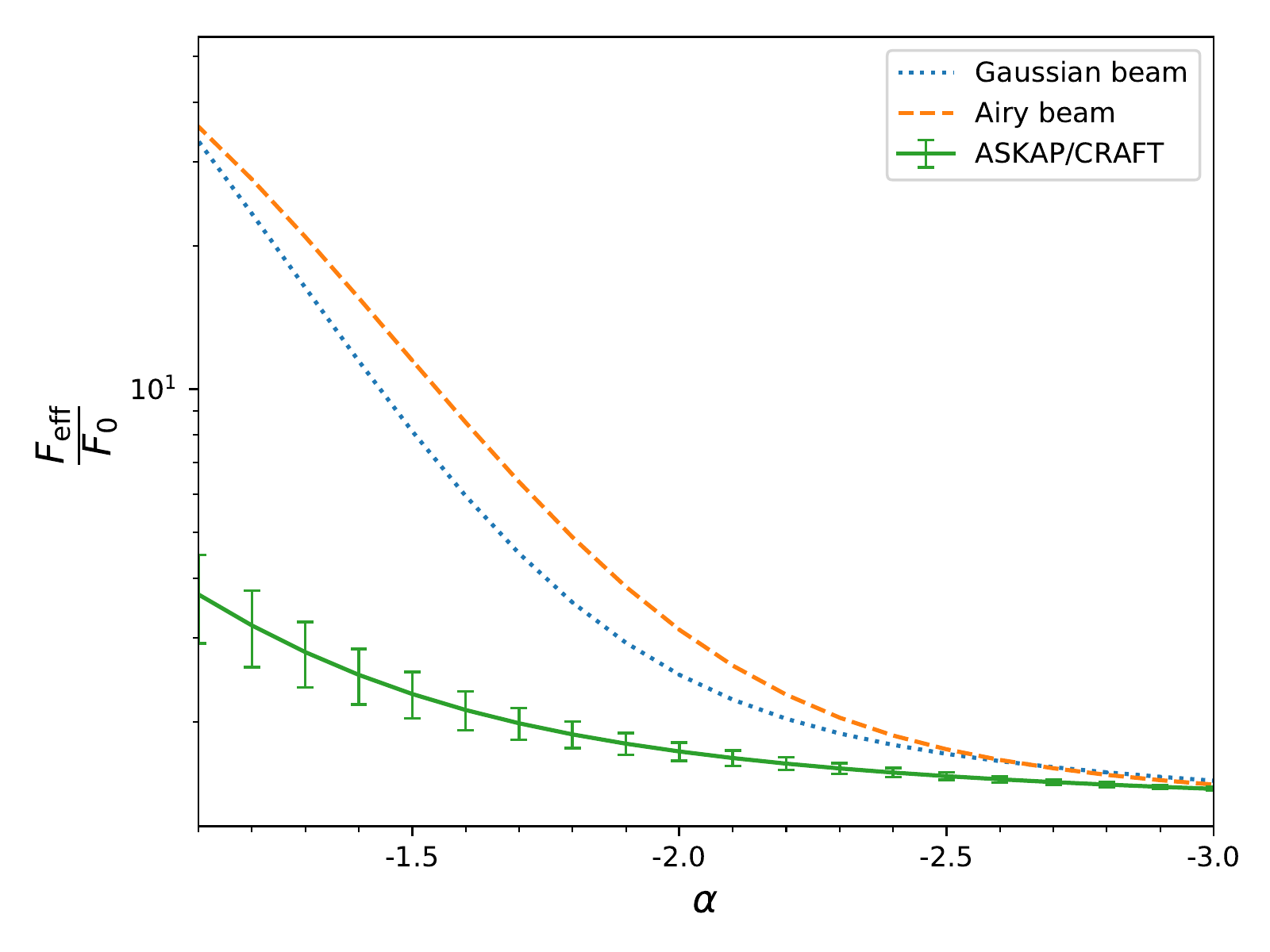} \includegraphics[width=0.49\textwidth]{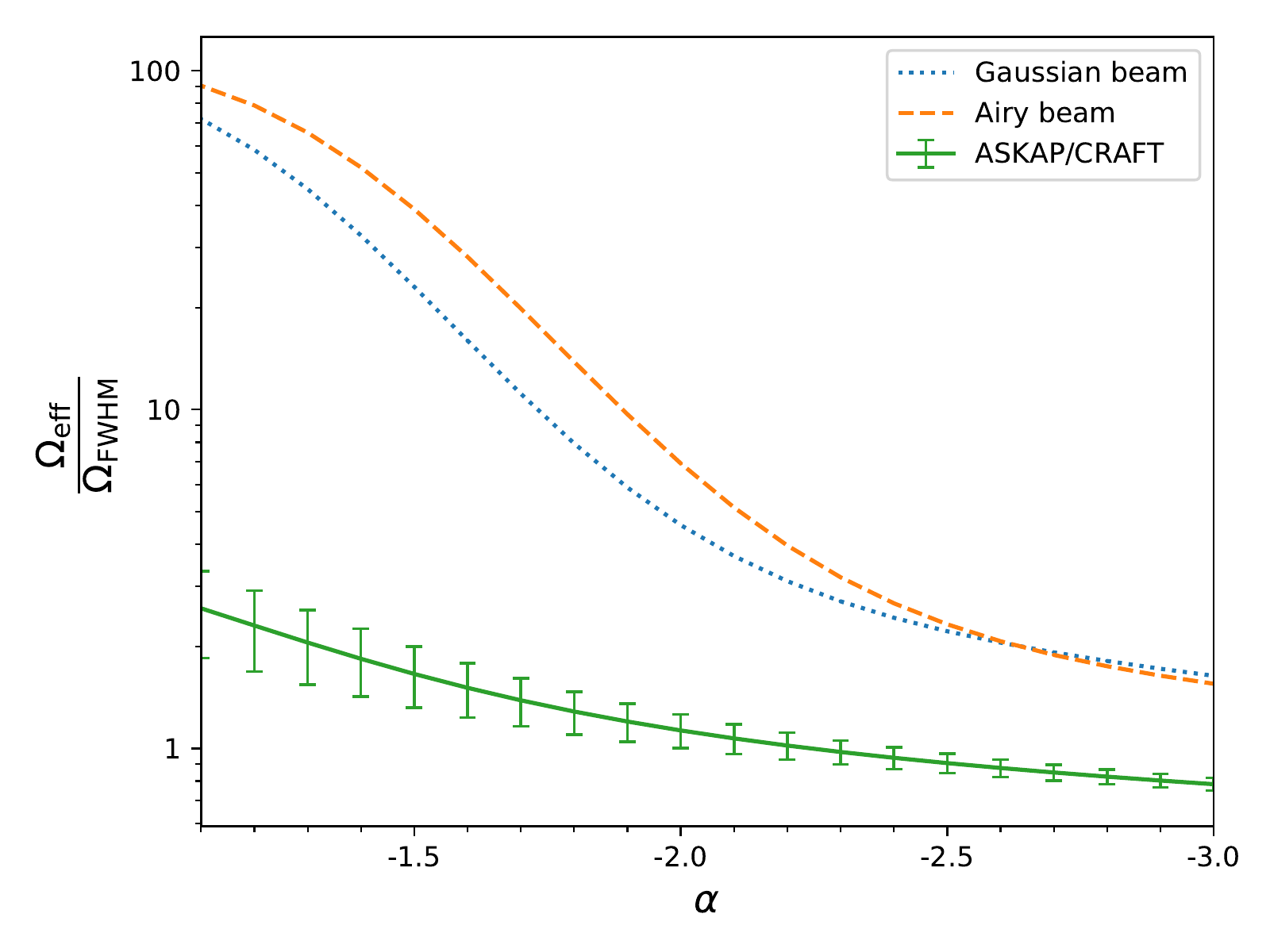} 
\caption{Effective observation parameters relative to their nominal values: effective fluence threshold $F_{\rm eff}/F_0$ (left) and effective survey area $\Omega_{\rm eff}/\Omega_{\rm FWHM}$ (right). CRAFT GL50 results are calculated from the mean of the best- and worst-case scenarios of $\Omega(B)$, with errors showing the systematic range corresponding to using each scenario. This is compared to results from single Airy and Gaussian beams. $F_0$ is relative to peak central beam sensitivity, while $\Omega_{\rm FWHM}$ is calculated as the beam full width half maximum for an Airy disk at central frequency.} \label{fig:feff_omegaeff}
\end{center}
\end{figure*}

For the CRAFT GL50 survey, both $F_{\rm eff}(\alpha)$ and $\Omega_{\rm eff}(\alpha)$, calculated as per equations (\ref{eq:fbar}) and (\ref{eq:eff}) respectively, are shown in Figure~\ref{fig:feff_omegaeff}. For comparison, results using single Airy and Gaussian beams are also shown.

In interpreting the result, note that $F_{\rm eff}$ will always be greater than the nominal sensitivity $F_0$ at beam centre. As $\alpha$ tends to $-1$, there are more FRBs with large amplitude. The effects of sidelobes become increasingly important, and $\Omega_{\rm eff}$ will be larger than quoted values (typically the area at full width half maximum). Conversely, as $\alpha$ tends to negative infinity, $F_{\rm eff}$ will tend to $F_0$, and $\Omega_{\rm eff}$ will tend to zero. In the case that $\alpha > -1$, observations will be so biased towards high-luminosity FRBs detected far from beam centre that any definition of $F_{\rm eff}$ will be physically meaningless. However, observations do not appear to be in this regime, given the current best fit to ASKAP/CRAFT data of $\alpha = -2.2$, and $\alpha=-1.2$ for Parkes \citep{2018arXiv181004357J}.

The effect of the overlapping beams used in CRAFT surveys is immediately apparent in Figure~\ref{fig:feff_omegaeff}. This flattens the dependence of $F_{\rm eff}$ and $\Omega_{\rm eff}$ on $\alpha$, whereas for Gaussian or Airy beams, these parameters vary by an order of magnitude between $\alpha=-1$ and $\alpha=-3$. The errors due to the uncertainty in the shape of the sidelobes of outer ASKAP beams become more important at low values of $\alpha$.

As discussed in the previous Section, the calculation is expected to become numerically unstable near $\alpha=-1$. However, the measurement of the beam over a finite patch of sky effectively cuts off the integration in equation~(\ref{eq:fbar}), achieving numerical stability at the cost of physical accuracy. To estimate the magnitude of this effect, we calculated $F_{\rm eff}(\alpha)$ for the closepack36 configuration using Gaussian beamshapes, both with and without limiting the integral to the $8.4\times8.4$\,deg region of the holography scans. The resulting error in $F_{\rm eff}$ was less than that due to the beamshape for $\alpha \le -1.1$. The calculations for Gaussian and Airy beams are exact.

The values from Figure~\ref{fig:feff_omegaeff} are compiled in Table~\ref{tbl:eff_data}. The relative increases in $F_{\rm eff}$ and $\Omega_{\rm eff}$ compared to nominal values are also given, which is particularly useful for scaling results from other experiments.

\begin{figure*}
\begin{center}
\includegraphics[width=0.7\textwidth]{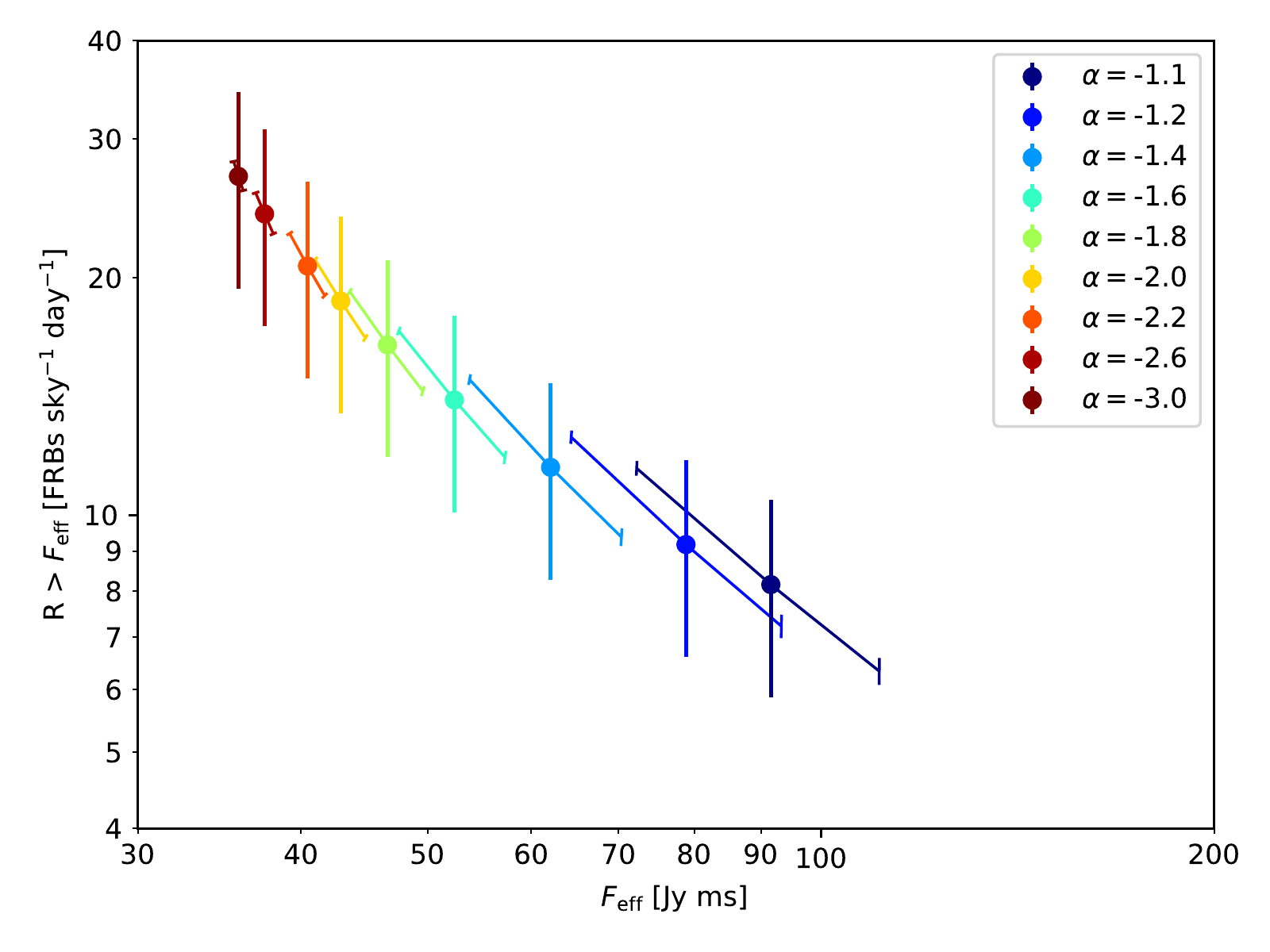}
\caption{Measured all-sky ($4 \pi$\,sr) rate of FRBs above the CRAFT GL50 effective fluence threshold, $F_{\rm eff}$, for different values of $\alpha$. Vertical error bars correspond to statistical (`stat') $1$\,$\sigma$ Poissonian errors from the 19 detections, while angled error bars correspond to systematic (`sys') errors in $F_{\rm eff}$ and $\Omega_{\rm eff}$ in Figure~\ref{fig:feff_omegaeff}.} \label{fig:craft_rates}
\end{center}
\end{figure*}

Using the 19 FRBs detected above threshold (FRB 171216 was detected only using a combination of beams in a non-standard analysis, and falls below the threshold calculated here), Figure~\ref{fig:craft_rates} shows the measured all-sky ($4 \pi$\,sr) rates as a function of both $\alpha$ and $F_{\rm eff}$.
For the range of $\alpha$ from $-1.5$ to $-2.7$ reported by \citet{2018arXiv181004357J} (68\% confidence), the rate varies between 12.7 and 24.9 FRBs\,sky$^{-1}$\,day$^{-1}$, above thresholds of 57 and 37\,Jy\,ms respectively. 
The error resulting from uncertainties in $F_{\rm eff}$ and $\Omega_{\rm eff}$ is approximately equal in impact to an uncertainty in $\alpha$ of $\pm 0.2$, and is the dominant source of error for $\alpha>-1.4$.

\section{DISCUSSION}
\label{sec:discussion}

The CRAFT GL50 survey has provided a large sample of events (20) with which to probe the nature of fast radio bursts. We have developed methods to account for the effects of beam- and antenna-dependent sensitivity, detection efficiency, and time-dependent noise effects during ASKAP's commissioning phase. In doing so, FRB 171216 had to be discarded for analysis purposes, since it was detected using a non-standard, and hence uncalibrated, method. Remaining uncertainties in these effects are comparable to the Poisson uncertainty due to the small number of detected FRBs. Since the latter will reduce with future detections, efforts should continue to quantify the ASKAP beam pattern, in particular the sidelobes of the outer beams.

The dependence of effective survey parameters $\Omega_{\rm eff}$ and $F_{\rm eff}$, and hence the measured FRB rate, on the integral source counts spectral index $\alpha$ dominates all other uncertainties. This highlights the importance of accurately modelling the effects discussed by \citet{2018MNRAS.474.1900M}, since ignoring them would constitute a systematic error greater than the uncertainties discussed above.

The estimated FRB rate varies greatly with the assumed spectral index of the integral source counts distribution. For a Euclidean power-law index of $\alpha=-1.5$, we find a rate of $12.7_{-2.2}^{+3.3}\,{\rm (sys)} \, \pm \, 3.6\,{\rm (stat)}$\,sky$^{-1}$\,day$^{-1}$ above a threshold of $56.6 \pm 6.3\,{\rm (sys)}$\,Jy\,ms, at the CRAFT time resolution of $1.2656$\,ms.

The studies performed by \citet{2016ApJ...830...75V} favoured a flat spectral index ($\alpha \gtrsim -1$). For $\alpha=-1.1$, we find a rate of $8.2_{-1.8}^{+3.3} \,{\rm (sys)}\, \pm 2.3\,{\rm (stat)}$ above a threshold of $92\pm19 (\rm sys)$. The rate is, not unexpectedly, much lower than that found with previous estimates at lower fluence thresholds \citep{2014ApJ...789L..26P,2016MNRAS.460L..30C,2018MNRAS.475.1427B}. Not only is the nominal CRAFT threshold higher, but these estimates did not include calculations of effective survey parameters, and should do so before sensible comparisons can be made. The methods of \citet{2017AJ....154..117L} are more appropriate: using Gaussian beamshapes, they find a best-fit $\alpha=-0.91 \pm 0.34$ (95\% C.I.), and a correspondingly lower FRB rate of $R=587^{+336}_{-315}$ above $1$\,Jy\,ms.

Using preliminary calibration parameters for the same CRAFT GL50 survey, \citet{craft_nature} have reported a measured rate of $37\pm8$\,sky$^{-1}$\,day$^{-1}$, and best-fit spectral index of $\alpha=-2.1^{+0.6}_{-0.5}$. This is quoted at a threshold of $46$\,Jy over $4$\,ms, ie.\ $26$\,Jy over $1.2656$\,ms. For the case of $\alpha=-2.1$, we find an effective threshold of $41.6\pm1.5\,{\rm (sys)}$\,Jy\,ms, and all-sky rate of $19.7_{-1.8}^{+2.2} \,{\rm (sys)}\, \pm 5.5\,{\rm (stat)}$\,sky$^{-1}$\,day$^{-1}$. The rate calculated here is lower due to several effects. Even for $\alpha=-3$ (where we find $26.9$\,sky$^{-1}$\,day$^{-1}$), the effective survey solid-angle, $\Omega_{\rm eff}$, is 20\% smaller than the $\Omega_{\rm FWHM}=35.9$\,deg$^2$ of 36 independent beams. However, it is 40\% larger than the effective field of view of $20$\,deg$^2$ used by \citet{craft_nature}. The effective observation time $T_{\rm eff}$ of $1060.8$ antenna days used in \citet{craft_nature} is slightly smaller than the $1108.9$ antenna days found here, due to a lower assumed efficiency (80\%, c.f.\ the 87\% found here) more than compensating for the data losses reducing the nominal observation time from $1326$ to $1274.6$ antenna days. In combination, the survey exposure used by \citet{craft_nature} is $5.1\cdot 10^5$\,deg$^2$\,hr, which is 27\% less than that of the $\alpha=-3$ value of $7.00\cdot 10^5$\,deg$^2$\,hr found here, and accounts for the lower all-sky rate found by this work.

We have not accounted for dependencies of FRB search sensitivity on their duration, DM, or frequency structure --- the values of $F_{\rm eff}$ quoted are with respect to an idealised pulse of $1.2656$\,ms duration. \citet{2015MNRAS.447.2852K} discuss several effects that reduce the search sensitivity when compared to such an idealisation. We aim to directly compare the search sensitivity of the CRAFT search algorithm, FREDDA, to other software packages in the near future.

We note that while the particulars of this method may be unique to FRB searches with ASKAP, the general methodology --- that of calculating survey exposure as a function of effective FRB sensitivity --- is requisite for any FRB survey, particularly those aiming to study the population statistics rather than specific details of particular bursts. Both \citet{2014ApJ...790..101S} and \citet{2016ApJ...830...75V} present detailed, although untested, beam models for the ALFA and multibeam receivers on Arecibo and Parkes respectively, but their effects on survey parameters are not calculated. We suggest that a calculation of at least the beam effects, as is performed here in Section~\ref{sec:beams}, be performed for all instruments searching for fast radio bursts.

Furthermore, the efficiency of FRB searches (e.g.\ lost time due to RFI) is generally not published. \citet{2018MNRAS.474.3847F} provide a detailed analysis of their classification algorithm for the ALFABURST search with Arecibo, finding an efficiency of 157/163 (96\%), but the loss prior to the classification stage is unclear. Regular monitoring of bright, stable pulsars appears to be a promising method for quantifying the efficiency, and provides a useful relative calibration of search sensitivity, as we demonstrate in Section~\ref{sec:pulsar_cal}.

Making meaningful comparisons between the results of different instruments will be impossible without similar calculations to those presented in this work being performed.

CRAFT searches for FRBs are on-going. Using the effective thresholds and solid angles from Figure~\ref{fig:feff_omegaeff} with the nominal observation time in antenna days (multiplied by the efficiency factor $0.87$) will provide a good estimate of survey sensitivity. As pulsar calibration observations are ongoing --- and CRAFT will soon use the commissioned antennas in commensal mode --- updated antenna sensitivities (c.f.\ Figure~\ref{fig:afit}) should be used with the corresponding exposure times to generate a sensitivity histogram (Figure~\ref{fig:sens_hist}. Convolving this with the appropriate beam-pattern (Figure~\ref{fig:beam_sens_hist}) will allow a new exposure histogram (Figure~\ref{fig:craft_exposure}) to be calculated. Combining data-sets implies simply adding the new exposure histogram to the old before calculating a new set of effective survey parameters as per Section~\ref{sec:effective}, while analyses treating the samples independently (e.g.\ as a function of Galactic latitude) will require two sets of these parameters.

\renewcommand{\arraystretch}{1.2}

\begin{table*}
\caption{Tabularised effective CRAFT survey parameters as a function of FRB source counts index $\alpha$. Parameters are the effective fluence threshold $F_{\rm eff}$, nominal threshold at beam centre $F_0$, effective solid angle $\Omega_{\rm eff}$, and nominal solid angle at full width half maximum $\Omega_{\rm FWHM}$. The effective rate $R$ is also calculated, corresponding to $19$ FRBs over $T_{\rm eff}=1108.9$ antenna days. Mean values and errors are systematic (`sys') and correspond to the means and errors from the exposure $E$ in Figure~\ref{fig:craft_exposure}. The exception is the statistical error (`stat') in the rate R corresponding to Poisson fluctuations in the number of observed FRBs. These are shown as the second, symmetric component of the error in R.}\label{tbl:eff_data}
\centering
\begin{tabular}{| c | c c c c c |  c c | c c |}
\hline\hline
& \multicolumn{5}{c|}{CRAFT}   & \multicolumn{2}{c|}{Gaussian} & \multicolumn{2}{c|}{Airy} \\
\hline
$\alpha$  
& $F_{\rm eff}$ & \multirow{2}{*}{$\mathlarger{\frac{F_{\rm eff}}{F_0}}$} & $\Omega_{\rm eff}$ & \multirow{2}{*}{$\mathlarger{\frac{\Omega_{\rm eff}}{\Omega_0}}$} & R 
& \multirow{2}{*}{$\mathlarger{\frac{F_{\rm eff}}{F_0}}$} & \multirow{2}{*}{$\mathlarger{\frac{\Omega_{\rm eff}}{\Omega_0}}$}
& \multirow{2}{*}{$\mathlarger{\frac{F_{\rm eff}}{F_0}}$} & \multirow{2}{*}{$\mathlarger{\frac{\Omega_{\rm eff}}{\Omega_0}}$} \\
& Jy ms & & deg$^2$ & &[sky$^{-1}$\,day$^{-1}$] &&&& \\
\hline
-1.1 &   91.5 $\pm$   19.3 &    3.7 $\pm$   0.78 &   86.6 $\pm$   25.0 &    2.4 $\pm$   0.68 & $    8.2_{-1.8}^{+3.3} \pm 2.3 $ &  33.14 &   72.0 &  35.54 &   90.3\\
-1.2 &   78.8 $\pm$   14.4 &    3.2 $\pm$   0.58 &   77.1 $\pm$   20.8 &    2.1 $\pm$   0.57 & $    9.2_{-1.9}^{+3.4} \pm 2.6 $ &  23.34 &   58.4 &  27.56 &   78.9\\
-1.3 &   69.3 $\pm$   10.9 &    2.8 $\pm$   0.44 &   68.7 $\pm$   17.1 &    1.9 $\pm$   0.47 & $   10.3_{-2.0}^{+3.4} \pm 2.9 $ &  16.31 &   44.7 &  20.86 &   65.6\\
-1.4 &   62.1 $\pm$    8.3 &    2.5 $\pm$   0.33 &   61.5 $\pm$   13.9 &    1.7 $\pm$   0.38 & $   11.5_{-2.1}^{+3.4} \pm 3.2 $ &  11.44 &   32.6 &  15.53 &   51.8\\
-1.5 &   56.6 $\pm$    6.3 &    2.3 $\pm$   0.26 &   55.5 $\pm$   11.3 &    1.5 $\pm$   0.31 & $   12.7_{-2.2}^{+3.3} \pm 3.6 $ &   8.15 &   23.0 &  11.48 &   39.0\\
-1.6 &   52.4 $\pm$    4.9 &    2.1 $\pm$   0.20 &   50.5 $\pm$    9.2 &    1.4 $\pm$   0.25 & $   14.0_{-2.2}^{+3.1} \pm 3.9 $ &   5.96 &   16.0 &   8.51 &   28.2\\
-1.7 &   49.1 $\pm$    3.8 &    2.0 $\pm$   0.15 &   46.4 $\pm$    7.6 &    1.3 $\pm$   0.21 & $   15.2_{-2.1}^{+3.0} \pm 4.3 $ &   4.51 &   11.1 &   6.38 &   19.9\\
-1.8 &   46.6 $\pm$    3.0 &    1.9 $\pm$   0.12 &   43.0 $\pm$    6.2 &    1.2 $\pm$   0.17 & $   16.4_{-2.1}^{+2.8} \pm 4.6 $ &   3.56 &    7.9 &   4.88 &   13.8\\
-1.9 &   44.5 $\pm$    2.4 &    1.8 $\pm$   0.10 &   40.2 $\pm$    5.2 &    1.1 $\pm$   0.14 & $   17.6_{-2.0}^{+2.6} \pm 4.9 $ &   2.93 &    5.9 &   3.84 &    9.7\\
-2.0 &   42.9 $\pm$    1.9 &    1.7 $\pm$   0.08 &   37.8 $\pm$    4.3 &    1.0 $\pm$   0.12 & $   18.7_{-1.9}^{+2.4} \pm 5.2 $ &   2.51 &    4.6 &   3.12 &    6.9\\
-2.1 &   41.6 $\pm$    1.5 &    1.7 $\pm$   0.06 &   35.8 $\pm$    3.6 &    1.0 $\pm$   0.10 & $   19.7_{-1.8}^{+2.2} \pm 5.5 $ &   2.23 &    3.7 &   2.62 &    5.1\\
-2.2 &   40.4 $\pm$    1.2 &    1.6 $\pm$   0.05 &   34.1 $\pm$    3.1 &    0.9 $\pm$   0.08 & $   20.7_{-1.7}^{+2.1} \pm 5.8 $ &   2.03 &    3.1 &   2.28 &    4.0\\
-2.3 &   39.5 $\pm$    1.0 &    1.6 $\pm$   0.04 &   32.7 $\pm$    2.7 &    0.9 $\pm$   0.07 & $   21.6_{-1.6}^{+1.9} \pm 6.1 $ &   1.89 &    2.7 &   2.04 &    3.2\\
-2.4 &   38.7 $\pm$    0.8 &    1.6 $\pm$   0.03 &   31.4 $\pm$    2.3 &    0.9 $\pm$   0.06 & $   22.5_{-1.5}^{+1.8} \pm 6.3 $ &   1.79 &    2.4 &   1.87 &    2.7\\
-2.5 &   38.1 $\pm$    0.7 &    1.5 $\pm$   0.03 &   30.3 $\pm$    2.0 &    0.8 $\pm$   0.05 & $   23.3_{-1.4}^{+1.6} \pm 6.5 $ &   1.71 &    2.2 &   1.75 &    2.3\\
-2.6 &   37.5 $\pm$    0.6 &    1.5 $\pm$   0.02 &   29.3 $\pm$    1.8 &    0.8 $\pm$   0.05 & $   24.1_{-1.4}^{+1.5} \pm 6.7 $ &   1.65 &    2.1 &   1.66 &    2.1\\
-2.7 &   37.0 $\pm$    0.5 &    1.5 $\pm$   0.02 &   28.4 $\pm$    1.6 &    0.8 $\pm$   0.04 & $   24.9_{-1.3}^{+1.4} \pm 7.0 $ &   1.61 &    1.9 &   1.60 &    1.9\\
-2.8 &   36.6 $\pm$    0.4 &    1.5 $\pm$   0.02 &   27.6 $\pm$    1.4 &    0.8 $\pm$   0.04 & $   25.6_{-1.2}^{+1.3} \pm 7.2 $ &   1.57 &    1.8 &   1.55 &    1.7\\
-2.9 &   36.2 $\pm$    0.4 &    1.5 $\pm$   0.01 &   26.9 $\pm$    1.2 &    0.7 $\pm$   0.03 & $   26.2_{-1.2}^{+1.3} \pm 7.3 $ &   1.53 &    1.7 &   1.51 &    1.6\\
-3.0 &   35.8 $\pm$    0.3 &    1.4 $\pm$   0.01 &   26.3 $\pm$    1.1 &    0.7 $\pm$   0.03 & $   26.9_{-1.1}^{+1.2} \pm 7.5 $ &   1.50 &    1.6 &   1.48 &    1.5\\
\hline
\hline
\end{tabular}
\end{table*}

\section{CONCLUSION}

We have derived the sensitivity and exposure of the CRAFT GL50 FRB survey with the Australian SKA Pathfinder (ASKAP). The ASKAP beam pattern, antenna- and time-dependent sensitivity, and loss of efficiency, have been accounted for. As such, not only do the 20 FRBs detected by the CRAFT GL50 survey constitute the largest single FRB sample, they are also the best-calibrated for use in source statistics studies. It is noteworthy that this is the first extensive astronomical survey performed using phased array feeds, and that this new technology has yielded such a detailed calibration.

Our methodology allows the calibration of future CRAFT FRB surveys with ASKAP, and points the way towards similar calculations being performed for other FRB searches. That one FRB must effectively be discarded due to the use of non-standard search methods highlights the importance of using well-defined thresholds for detection.

We have for the first time calculated the dependence of effective survey threshold and exposure on the spectral index of the FRB integral source counts distribution. This includes a full analysis of systematic uncertainties (`sys'). The closer beam spacing used for the CRAFT GL50 survey with ASKAP results in a smaller range of variation than expected for other beam footprints. Nonetheless, we find that the variation in these parameters is greater than the statistical uncertainty (`stat') in the FRB rate due to the small number of detected events.

Using effective rather than nominal survey parameters, the rate of $37$\,sky$^{-1}$\,day$^{-1}$ found by \citet{craft_nature} is reduced to $12.7_{-2.2}^{+3.3}\,{\rm (sys)} \, \pm \, 3.6\,{\rm (stat)}$\,sky$^{-1}$\,day$^{-1}$ at the Euclidian expectation of $\alpha=-1.5$ for the source-counts index. At the best-fit value of the source-counts index ($\alpha=-2.2$; \citet{2018arXiv181004357J}), the rate is $20.7_{-1.7}^{+2.1} \,{\rm (sys)}\, \pm 5.8\,{\rm (stat)}$\,sky$^{-1}$\,day$^{-1}$. The corresponding thresholds are $56.6 \pm 6.3\,{\rm (sys)}$ and $40.4 \pm 1.2\,{\rm (sys)}$\,Jy\,ms respectively, at the CRAFT time resolution of $1.2656$\,ms.

The increased precision of this calculation highlights the remaining unknowns. In particular, the DM-dependence of survey sensitivity should also be quantified. We encourage the authors of other FRB surveys to perform similar calculations for their instruments and methods.

\begin{acknowledgements}
We thank N.\ Tejos for comments on the manuscript, and the MWA principle scientist, R.\ Wayth, for access to the Galaxy supercomputer GPU cluster. R.M.S. and S.O. acknowledge Australian Research Council grant FL150100148. R.M.S. also acknowledges support through grant CE170100004. Parts of this research  were conducted by the Australian Research Council Centres of Excellence for All Sky Astrophysics (CAASTRO, CE110001020) and All Sky Astrophysics in 3 Dimensions (ASTRO3D, CE170100013). This research was also supported by the Australian Research Council through grant DP18010085. The Australian SKA Pathfinder and Parkes radio telescopes are part of the Australia Telescope National Facility which is managed by CSIRO. Operation of ASKAP is funded by the Australian Government with support from the National Collaborative Research Infrastructure Strategy. ASKAP uses the resources of the Pawsey Supercomputing Centre. Establishment of ASKAP, the Murchison Radio-astronomy Observatory and the Pawsey Supercomputing Centre are initiatives of the Australian Government, with support from the Government of Western Australia and the Science and Industry Endowment Fund. We acknowledge the Wajarri Yamatji people as the traditional owners of the Observatory site.
\end{acknowledgements}

\bibliographystyle{pasa-mnras}
\bibliography{bibtex_entries.bib}

\end{document}